\documentclass[10pt]{iopart}

\usepackage[varg]{txfonts}
\usepackage{bm}
\usepackage{graphicx}
\usepackage{latexsym}
\usepackage{supertabular}
\usepackage{iopams}
\usepackage{hyperref}
\usepackage{color}

\def\rd{\mathrm{d}}

\def\Ry{\,\mathrm{Ry}}
\def\kelvin{\,\mathrm{K}}

\begin{document}

\title[Electron-impact excitation of N$^{3+}$ using the BSR method.]
      {Electron-impact excitation of N$^{3+}$ using the B-spline R-matrix method: 
       Importance of the target structure description and the size of the 
       close-coupling expansion.}
\author{L Fern\'{a}ndez-Menchero, O Zatsarinny and K Bartschat}
\address{Department of Physics and Astronomy, Drake University, Des Moines IA, 50311, USA}

\begin{abstract}

There are major discrepancies between recent ICFT (Intermediate Coupling
Frame Transformation) and DARC (Dirac Atomic R-matrix Code) calculations
(Fern\'andez-Menchero {\em et al.}~2014, {\em Astron.\ Astroph.}~{\bf 566} A104,
 Aggarwal {\em et al.}~2016 {\em Mon.\ Not.\ R.~Astr.\ Soc.}~{\bf 461} 3997)
regarding electron-impact excitation rates for transitions in several Be-like ions, as
well as claims that the DARC calculations are much more accurate and the ICFT results might even be wrong.
To identify possible reasons for these discrepancies and to estimate the accuracy 
of the various results, we carried out independent B-Spline R-Matrix (BSR) calculations
for electron-impact excitation of the Be-like ion N$^{3+}$.
Our close-coupling expansions contain the same target states (238 levels overall) 
as the previous ICFT and DARC calculations, 
but the representation of the target wave functions is completely different.
We find close agreement among all calculations for the strong transitions
between low-lying states, whereas there remain serious discrepancies for the weak transitions
as well as for transitions to highly excited states.
The differences in the final results for the collision strengths are mainly due to differences
in the structure description, specifically the inclusion of correlation effects, rather 
than the treatment of relativistic effects or problems with the validity of the 
three methods to describe the collision. 
Hence there is no indication that one approach is superior to another,  
until the convergence of both the target configuration 
and the close-coupling expansions have been fully established.

\end{abstract}

\pacs{34.50.Fa,52.20.Fs,95.30.Ky}

\submitto{\JPB}
\maketitle
\ioptwocol

\section{Introduction}
\label{sec:introduction}

Accurate and reliable electron-impact excitation data are required for
the modeling and spectroscopic diagnostics of various non-equilibrium
astrophysical and laboratory plasmas. 
Emission lines from beryllium-like ions (in particular N$^{3+}$) are used for 
electron density and temperature diagnostics for a variety of emission sources. 
Many references and examples are given in previous publications on Be-like 
ions~\cite{fernandez-menchero2014a,aggarwal2016}.
For the processes of interest here, the Be-like ions can be modeled well as two-electron systems with a frozen 
$\mathrm{1s^2}$ core.
Despite the simple structure of these ions, 
the electron-impact collision strengths from recent 
calculations~\cite{fernandez-menchero2014a,aggarwal2016} differ considerably.
This calls for additional analysis of the methods used to obtain these results, as well as the existing data.

Many formulations and the associated computer codes have been developed to treat electron scattering from atoms and ions.
The most frequently applied {\it general} approach is the {\hbox{R-matrix} } method 
developed by Burke and collaborators over the past decades.  
The basic ideas, as well as numerous applications and extensions to other fields 
and processes, are described in~\cite{burke2011}.
The suite of Belfast {\hbox{R-matrix}} codes~\cite{berrington1995} 
(see, for example, \cite{Rmax} for updates) is, 
within the usual computational restraints, applicable to the calculation
of electron (and photon) induced processes for any atomic or ionic target.
Relativistic corrections, which are only expected to be important for heavy targets,
are often included through the Breit-Pauli approximation.
An alternative and, in principle, more satisfactory approach is to use the Dirac-Coulomb hamiltonian,
as implemented in the full-relativistic 
Dirac Atomic \hbox{R-matrix} Code (DARC)~\cite{DARC}.  However, even
for the significantly heavier (than N$^{3+}$) target Fe$^{14+}$ only small differences between
Dirac-Coulomb and Breit-Pauli results were found in a detailed comparison study~\cite{Berrington2005},
provided the target structure description was similar in the two calculations.  Hence, it is
very unlikely that the semi-relativistic Breit-Pauli approach is insufficient for a light system such as N$^{3+}$
regarding the treatment of relativistic effects in generating data for practical modeling applications. 

Like most standard implementations of atomic structure and collision codes, 
the above-mentioned suites of programs employ a fixed set of mutually orthogonal one-electron orbitals 
for {\it all} states included in the expansion of the total wave function.  
In addition, the orbitals employed for the description of the projectile 
are made orthogonal to those used in the target description.
These choices are due to technical rather than physical reasons.  
(Note that only the multi-electron wave functions of the same symmetry and 
different total energies have to be orthogonal, 
but not the individual one-electron orbitals.)  
A significant advantage of the above choice is a simplification in setting up 
the hamiltonian matrix.  
On the other hand, these restrictions limit the flexibility in the 
description of the target structure.

The present calculations were carried out with a parallelized version 
of the \hbox{B-spline} Breit-Pauli \hbox{R-matrix} 
(BSR) code~\cite{zatsarinny2006}.
A distinct feature of the method (see~\cite{zatsarinny2013} for a comprehensive overview)
is the use of term-dependent
non-orthogonal orbital sets in the description of the target states.
This allows us to optimize the wave functions for different states 
independently, thereby resulting in a more accurate target description than those used in 
previous collision calculations.
The BSR code has been primarily applied to electron scattering from neutral 
atoms~\cite{zatsarinny2013},
where electron correlation effects are most important and an accurate 
representation of the target states becomes a primary task.
It has also been  applied to electron collisions with multi-charged ions, 
including Fe$^{6+}$~\cite{tayal2014} and Fe$^{7+}$~\cite{tayal2011}. 
Both of these ions have a complicated structure with open 3p and 3d sub\-shells, 
and an accurate representation of the target states even for these highly-charged
ions is not a trivial task. 
The distinct feature of the above calculations was a careful analysis of the 
convergence of the CI expansions for the target states.

The semi-relativistic (Breit-Pauli) or full-relativistic (Dirac-Coulomb) 
approaches require much more computational effort than non-relativistic 
calculations.
An alternative to a full Breit-Pauli \hbox{R-matrix}  calculation for electron-impact
excitation is the algebraic transformation of the scattering~($\bm{S}$) or reactance~($\bm{K}$) matrices, 
first calculated in pure $LS$-coupling,
to a relativistic coupling scheme.
In particular, Griffin {\em et al.}~\cite{griffin1998} introduced the 
intermediate-coupling frame transformation (ICFT) method, which employs multi-channel 
quantum defect theory (MQDT).
Even though the spin-orbit interaction between the colliding electron 
and the target is neglected in the ICFT approach, it turned out to be a very accurate approximation 
to the direct Breit-Pauli approach.
Numerous calculations~\cite{fernandez-menchero2015b, delzanna2015b} showed that the 
differences observed between the ICFT and the other \hbox{R-matrix}  results are well within 
the uncertainties to be expected due to the use of different target descriptions or
different close-coupling (CC) expansions for the scattering electron. 
At the same time, the ICFT method is computationally much more effective and 
has been applied in systematic calculations for entire isoelectronic sequences.

Fern\'andez-Menchero {\em et al.}~\cite{fernandez-menchero2014a} 
performed extensive \hbox{R-matrix} ICFT calculations for the 
$\mathrm{Be}$-like isoelectronic sequence up to $\mathrm{Kr}^{32+}$, 
including $\mathrm{N}^{3+}$.
For the latter, their scattering model included 238 intermediate-coupling (IC) levels.
Later, Aggarwal and coworkers published 98-level \hbox{R-matrix}
calculations obtained with the DARC code for the $\mathrm{Be}$-like 
ion $\mathrm{Al}^{9+}$~\cite{aggarwal2015a} and a 166-level calculation 
for $\mathrm{C}^{2+}$~\cite{aggarwal2015b}.
The differences in the obtained collision rates compared to the ICFT 
results of \cite{fernandez-menchero2014a} led 
Aggarwal {\em et al.}\ to cast doubt on the validity of the ICFT approach in general and/or the 
practical implementation.  
Fernandez-Menchero~\cite{fernandez-menchero2015b}, however, showed that the different results were 
simply due to variations in the description of 
the target structure among the calculations.
Similar criticism of ICFT was put forward by Aggarwal {\em et al.}\ regarding the 
ion $\mathrm{Fe}^{13+}$ \cite{aggarwal2014c}, but their arguments were 
again rebutted by Del~Zanna {\em et al}.~\cite{delzanna2015b}.

Recently, new calculations for the 
$\mathrm{Be}$-like ion $\mathrm{N}^{3+}$ appeared~\cite{aggarwal2016}, this time 
also obtained from a 238-level expansion with the
same set of configurations as used in~\cite{fernandez-menchero2014a}.
Since, once again, differences were found, the question remains whether these are 
mainly due to the treatment of the collision part of the problem or the still different
structure descriptions of the 238 target states.

The principal motivation for the present work, therefore, was to shed more light 
on the ongoing discussion by performing an independent calculation for 
electron collisions with $\mathrm{N}^{3+}$.  We use the BSR approach, 
which employs entirely different numerical methods for both the target states and the 
collision calculations.
For a proper comparison of the results, we select the same set of target states in 
the CC expansion as in previous ICFT~\cite{fernandez-menchero2014a} and 
DARC~\cite{aggarwal2016} calculations.
This allows us to directly draw conclusions regarding the sensitivity of the predictions to the target structure 
description in the case of interest.  Furthermore, we note that the BSR approach is expected to
have the most accurate representation of the target states, due to its flexibility in 
employing term-dependent, individually optimized, and hence non-orthogonal one-electron  
orbitals.

This manuscript is organized as follows.
In Sect.~\ref{sec:structure} we describe the structure calculations.  This is followed with 
some details about the close-coupling expansion in Sect.~\ref{sec:scattering}, where we also 
summarize the  way to obtain the collision strength~$\Omega$ and
the effective collision strength~$\Upsilon$.  
Our results are presented and discussed in Sect.~\ref{sec:results}, and the 
conclusions are summarized in Sect.~\ref{sec:conclusions}.
Unless specified otherwise, atomic units are used throughout.

\section{Structure Model}
\label{sec:structure}

The target states of $\mathrm{N}^{3+}$ in the present calculations were generated 
using the \hbox{B-spline} box-based close-coupling 
method~\cite{zatsarinny2002,zatsarinny2009}.
Relativistic effects were accounted for at the level of the Breit-Pauli approximation in
the intermediate-coupling scheme. 
Specifically, the structure of the multichannel target expansions 
for states with given total electronic angular momentum~$J$ and parity was chosen as
\begin{eqnarray}
   \Phi^{2lnl,J} & = & \sum_{nl,LS} a_{nl}^{LSJ} \{\phi(\mathrm{2s})\,P_{nl}\}^{LS} +
                       \sum_{nl,LS} b_{nl}^{LSJ} \{\phi(\mathrm{2p})\,P_{nl}\}^{LS} \nonumber \\
            & & +  \sum_{nl,LS} c_{nl}^{LSJ} \{\phi(\mathrm{3s})\,P_{nl}\}^{LS} +
                   \sum_{nl,LS} d_{nl}^{LSJ} \{\phi(\mathrm{3p})\,P_{nl}\}^{LS}  \nonumber \\
            & & +  \sum_{nl,LS} e_{nl}^{LSJ} \{\phi(\mathrm{3d})\,P_{nl}\}^{LS} \, .
\label{eq:Phi}
\end{eqnarray}
Here $\phi$ denotes the Hartree-Fock (HF) functions of the three-electron $\mathrm{1s^2}nl$  
ionic states, with the core $\mathrm{1s}$ orbital frozen; 
$P_{nl}$ denotes the wave function of the outer valence electron, 
and $\{\,\}$ stands for an antisymmetrized product of functions, conserving
the total orbital angular momentum $L$ and the spin $S$.
The above expansion is similar to the one used in our recent calculations
for electron scattering from neutral Be~\cite{zatsarinny2016}.
Usually we employ separate multi-configuration Hartree-Fock (MCHF) expansions for a more accurate representation
of the low-lying states with equivalent electrons, in particular $\mathrm{2s^2}$ and $\mathrm{2p^2}$ 
in the present case.
However, we found that these states are already represented accurately by the 
expansions~(\ref{eq:Phi}) alone, and hence the expansions for $\mathrm{N}^{3+}$ could be 
simplified in comparison to those for neutral Be. 
Bearing in mind that the final multi-configuration expansions still need to be dealt 
with in the subsequent collision calculation with one more electron to be coupled in,
such simplification is very helpful from a computational point of view.

The unknown radial functions $P_{nl}(r)$ for the outer valence
electron were expanded in a \hbox{B-spline} basis
\begin{equation}
   P_{nl}(r)\ =\ \sum_k c_k\,B_k(r) \,,
\label{eq:PBSR}
\end{equation}
where $B_k(r)$ denotes an individual B-spline. 
The coefficients $c_k$ for each valence orbital $P_{nl}(r)$, as well as the
various coefficients in Eq.~(\ref{eq:Phi})
were obtained by diagonalizing the $N$-electron target hamiltonian,
including all one-electron Breit-Pauli corrections.
The corresponding equations were solved subject to the condition
that the wave functions vanish at the boundary.
The number of spectroscopic bound states that can be generated in the above 
scheme depends on the \hbox{B-spline} box radius. 
In the present calculations, this radius was set to $40\,a_0$, 
where $a_0 = 0.529 \times 10^{-10}\,$m is the Bohr radius.
The expansion~(\ref{eq:PBSR}) included 134 \hbox{B-splines} of order~8,
with the step~$h$ between the knots varying from $0.125$ to $0.4\,a_0$.
This allowed us to obtain accurate descriptions of the $\mathrm{N}^{3+}$ states
with principal quantum number for the valence electron up to $n=7$.
Note that the above \hbox{B-spline} bound-state close-coupling calculations 
generate different non-orthogonal sets of one-electron orbitals for each target state, 
and hence their subsequent use is somewhat complicated. 
Our configuration expansions 
contained between 10 and 50 configurations for each state and could 
still be used in the collision calculations with moderate computational resources.

\begin{table*}
\caption{The lowest 40 energy Levels of $\mathrm{N}^{3+}$.}
\label{tab:energies}
\begin{indented}
\item[\hspace{0.1\textwidth}]\begin{tabular}{@{}rllrr@{\,(}r@{)\quad}r@{\,(}r@{)\quad}r@{\,(}r@{)\quad}r}
\br
 $i$   &              Conf &                Level &        NIST &        BSR & $  \% $ &      GRASP & $  \% $ &         AS & $  \% $ & $\tau$          \\
\mr
 $  1$ & $\mathrm{2s^2}  $ & $\mathrm{^1S_0^{e}}$ & $     0.0$  & $     0.0$ & $   - $ & $     0.0$ & $   - $ & $     0.0$ & $   - $ & stable          \\
 $  2$ & $\mathrm{2s\,2p}$ & $\mathrm{^3P_0^{o}}$ & $ 67209.2$  & $ 67388.1$ & $  0.3$ & $ 67811.1$ & $  0.9$ & $ 68401.0$ & $  1.8$ & stable          \\
 $  3$ & $\mathrm{2s\,2p}$ & $\mathrm{^3P_1^{o}}$ & $ 67272.3$  & $ 67458.7$ & $  0.3$ & $ 67870.3$ & $  0.9$ & $ 68485.0$ & $  1.8$ & $ 2.070\,[ -3]$ \\
 $  4$ & $\mathrm{2s\,2p}$ & $\mathrm{^3P_2^{o}}$ & $ 67416.3$  & $ 67600.5$ & $  0.3$ & $ 68008.6$ & $  0.9$ & $ 68655.0$ & $  1.8$ & $ 8.590\,[ +1]$ \\
 $  5$ & $\mathrm{2s\,2p}$ & $\mathrm{^1P_1^{o}}$ & $130693.9$  & $132950.2$ & $  1.7$ & $138128.6$ & $  5.7$ & $138278.0$ & $  5.8$ & $ 4.060\,[-10]$ \\
 $  6$ & $\mathrm{2p^2}  $ & $\mathrm{^3P_0^{e}}$ & $175535.4$  & $177326.3$ & $  1.0$ & $178454.8$ & $  1.7$ & $179729.0$ & $  2.4$ & $ 5.360\,[-10]$ \\
 $  7$ & $\mathrm{2p^2}  $ & $\mathrm{^3P_1^{e}}$ & $175608.1$  & $177396.9$ & $  1.0$ & $178524.0$ & $  1.7$ & $179813.0$ & $  2.4$ & $ 5.360\,[-10]$ \\
 $  8$ & $\mathrm{2p^2}  $ & $\mathrm{^3P_2^{e}}$ & $175732.9$  & $177536.5$ & $  1.0$ & $178639.2$ & $  1.7$ & $179979.0$ & $  2.4$ & $ 5.350\,[-10]$ \\
 $  9$ & $\mathrm{2p^2}  $ & $\mathrm{^1D_2^{e}}$ & $188882.5$  & $190712.2$ & $  1.0$ & $195946.9$ & $  3.7$ & $197490.0$ & $  4.6$ & $ 4.380\,[ -9]$ \\
 $ 10$ & $\mathrm{2p^2}  $ & $\mathrm{^1S_0^{e}}$ & $235369.3$  & $241108.7$ & $  2.4$ & $247825.3$ & $  5.3$ & $248501.0$ & $  5.6$ & $ 3.020\,[-10]$ \\
 $ 11$ & $\mathrm{2s\,3s}$ & $\mathrm{^3S_1^{e}}$ & $377284.8$  & $376494.3$ & $ -0.2$ & $375937.0$ & $ -0.4$ & $374017.0$ & $ -0.9$ & $ 1.120\,[-10]$ \\
 $ 12$ & $\mathrm{2s\,3s}$ & $\mathrm{^1S_0^{e}}$ & $388854.6$  & $388219.7$ & $ -0.2$ & $387760.1$ & $ -0.3$ & $386213.0$ & $ -0.7$ & $ 4.090\,[-10]$ \\
 $ 13$ & $\mathrm{2s\,3p}$ & $\mathrm{^1P_1^{o}}$ & $404522.4$  & $403910.7$ & $ -0.2$ & $404057.2$ & $ -0.1$ & $402317.0$ & $ -0.5$ & $ 7.760\,[-11]$ \\
 $ 14$ & $\mathrm{2s\,3p}$ & $\mathrm{^3P_0^{o}}$ & $405971.6$  & $405292.5$ & $ -0.2$ & $404875.8$ & $ -0.3$ & $403110.0$ & $ -0.7$ & $ 8.290\,[ -9]$ \\
 $ 15$ & $\mathrm{2s\,3p}$ & $\mathrm{^3P_1^{o}}$ & $405987.5$  & $405310.3$ & $ -0.2$ & $404891.2$ & $ -0.3$ & $403128.0$ & $ -0.7$ & $ 8.040\,[ -9]$ \\
 $ 16$ & $\mathrm{2s\,3p}$ & $\mathrm{^3P_2^{o}}$ & $406022.8$  & $405344.8$ & $ -0.2$ & $404925.2$ & $ -0.3$ & $403161.0$ & $ -0.7$ & $ 8.230\,[ -9]$ \\
 $ 17$ & $\mathrm{2s\,3d}$ & $\mathrm{^3D_1^{e}}$ & $420045.8$  & $419102.3$ & $ -0.2$ & $419130.7$ & $ -0.2$ & $417332.0$ & $ -0.6$ & $ 3.300\,[-11]$ \\
 $ 18$ & $\mathrm{2s\,3d}$ & $\mathrm{^3D_2^{e}}$ & $420049.6$  & $419109.5$ & $ -0.2$ & $419134.0$ & $ -0.2$ & $417339.0$ & $ -0.6$ & $ 3.300\,[-11]$ \\
 $ 19$ & $\mathrm{2s\,3d}$ & $\mathrm{^3D_3^{e}}$ & $420058.0$  & $419120.3$ & $ -0.2$ & $419142.8$ & $ -0.2$ & $417350.0$ & $ -0.6$ & $ 3.300\,[-11]$ \\
 $ 20$ & $\mathrm{2s\,3d}$ & $\mathrm{^1D_2^{e}}$ & $429159.6$  & $428645.9$ & $ -0.1$ & $430447.9$ & $  0.3$ & $428315.0$ & $ -0.2$ & $ 5.420\,[-11]$ \\
 $ 21$ & $\mathrm{2p\,3s}$ & $\mathrm{^3P_0^{o}}$ & $465291.8$  & $464987.9$ & $ -0.1$ & $465023.9$ & $ -0.1$ & $463467.0$ & $ -0.4$ & $ 1.460\,[-10]$ \\
 $ 22$ & $\mathrm{2p\,3s}$ & $\mathrm{^3P_1^{o}}$ & $465371.0$  & $465066.2$ & $ -0.1$ & $465101.9$ & $ -0.1$ & $463552.0$ & $ -0.4$ & $ 1.460\,[-10]$ \\
 $ 23$ & $\mathrm{2p\,3s}$ & $\mathrm{^3P_2^{o}}$ & $465536.6$  & $465226.9$ & $ -0.1$ & $465263.2$ & $ -0.1$ & $463725.0$ & $ -0.4$ & $ 1.450\,[-10]$ \\
 $ 24$ & $\mathrm{2p\,3s}$ & $\mathrm{^1P_1^{o}}$ & $473029.3$  & $472983.1$ & $ -0.0$ & $474725.8$ & $  0.4$ & $473301.0$ & $  0.1$ & $ 1.200\,[-10]$ \\
 $ 25$ & $\mathrm{2p\,3p}$ & $\mathrm{^1P_1^{e}}$ & $480884.2$  & $480661.1$ & $ -0.0$ & $480548.5$ & $ -0.1$ & $478848.0$ & $ -0.4$ & $ 1.290\,[-10]$ \\
 $ 26$ & $\mathrm{2p\,3p}$ & $\mathrm{^3D_1^{e}}$ & $484498.2$  & $484323.7$ & $ -0.0$ & $484764.6$ & $  0.1$ & $482889.0$ & $ -0.3$ & $ 2.750\,[-10]$ \\
 $ 27$ & $\mathrm{2p\,3p}$ & $\mathrm{^3D_2^{e}}$ & $484594.9$  & $484419.2$ & $ -0.0$ & $484861.2$ & $  0.1$ & $482988.0$ & $ -0.3$ & $ 2.760\,[-10]$ \\
 $ 28$ & $\mathrm{2p\,3p}$ & $\mathrm{^3D_3^{e}}$ & $484746.2$  & $484567.3$ & $ -0.0$ & $485010.4$ & $  0.1$ & $483142.0$ & $ -0.3$ & $ 2.750\,[-10]$ \\
 $ 29$ & $\mathrm{2p\,3p}$ & $\mathrm{^3S_1^{e}}$ & $487607.4$  & $487167.3$ & $ -0.1$ & $487384.0$ & $ -0.0$ & $485622.0$ & $ -0.4$ & $ 9.980\,[-11]$ \\
 $ 30$ & $\mathrm{2s\,4s}$ & $\mathrm{^1S_0^{e}}$ & $      - $  & $494216.6$ & $   - $ & $493653.3$ & $   - $ & $492046.0$ & $   - $ & $ 5.410\,[-10]$ \\
 $ 31$ & $\mathrm{2p\,3p}$ & $\mathrm{^3P_0^{e}}$ & $494253.1$  & $494474.0$ & $  0.0$ & $494289.8$ & $  0.0$ & $493059.0$ & $ -0.2$ & $ 1.920\,[-10]$ \\
 $ 32$ & $\mathrm{2p\,3p}$ & $\mathrm{^3P_1^{e}}$ & $494309.2$  & $494511.4$ & $  0.0$ & $494334.8$ & $  0.0$ & $493110.0$ & $ -0.2$ & $ 1.860\,[-10]$ \\
 $ 33$ & $\mathrm{2p\,3p}$ & $\mathrm{^3P_2^{e}}$ & $494402.0$  & $494607.6$ & $  0.0$ & $494422.6$ & $  0.0$ & $493217.0$ & $ -0.2$ & $ 1.860\,[-10]$ \\
 $ 34$ & $\mathrm{2p\,3d}$ & $\mathrm{^3F_2^{o}}$ & $495406.2$  & $494807.6$ & $ -0.1$ & $495606.6$ & $  0.0$ & $494003.0$ & $ -0.3$ & $ 1.830\,[ -8]$ \\
 $ 35$ & $\mathrm{2p\,3d}$ & $\mathrm{^3F_3^{o}}$ & $495482.6$  & $494884.7$ & $ -0.1$ & $495680.2$ & $  0.0$ & $494087.0$ & $ -0.3$ & $ 2.040\,[ -8]$ \\
 $ 36$ & $\mathrm{2p\,3d}$ & $\mathrm{^3F_4^{o}}$ & $495585.7$  & $494986.6$ & $ -0.1$ & $495777.8$ & $  0.0$ & $494196.0$ & $ -0.3$ & $ 2.110\,[ -8]$ \\
 $ 37$ & $\mathrm{2s\,4s}$ & $\mathrm{^3S_1^{e}}$ & $498045.5$  & $497406.9$ & $ -0.1$ & $497108.9$ & $ -0.2$ & $495026.0$ & $ -0.6$ & $ 2.320\,[ -9]$ \\
 $ 38$ & $\mathrm{2p\,3d}$ & $\mathrm{^1D_2^{o}}$ & $498310.3$  & $497915.1$ & $ -0.1$ & $497699.3$ & $ -0.1$ & $496164.0$ & $ -0.4$ & $ 7.200\,[-11]$ \\
 $ 39$ & $\mathrm{2p\,3p}$ & $\mathrm{^1D_2^{e}}$ & $499705.9$  & $499803.8$ & $  0.0$ & $501828.7$ & $  0.4$ & $500670.0$ & $  0.2$ & $ 1.030\,[-10]$ \\
 $ 40$ & $\mathrm{2s\,4p}$ & $\mathrm{^3P_2^{o}}$ & $503680.4$  & $502965.3$ & $ -0.1$ & $502495.9$ & $ -0.2$ & $500770.0$ & $ -0.6$ & $ 1.820\,[-10]$ \\
\br
\end{tabular}
\item[]Key: $i$, level index; Conf, Level: Configuration and multiplet, largest weight;
BSR: calculated excitation energy relative to the ground state, present work;
NIST: recommended value from the NIST data base \cite{moore1993};
AS: work \cite{fernandez-menchero2014a}.
GRASP: work \cite{aggarwal2016};
$\%$: relative deviation with respect to the recommended values;
$\tau$: life-time (in seconds), present work.
All energies are given in $\mathrm{cm}^{-1}$.
$A\,[B]$ denotes $A \times 10^{B}$.

\end{indented}

\end{table*}

Standard \hbox{R-matrix} calculations employ CI expansions of the target states
based on a single set of orthogonal one-electron radial functions~\cite{burke2011}.
The ICFT calculations for Be-like ions~\cite{fernandez-menchero2014a}, for example, 
used atomic levels generated by {\tt AUTOSTRUCTURE} (AS)~\cite{badnell2011b},
with the $\mathrm{1s}$ to $\mathrm{7d}$ orbitals 
obtained from a Thomas-Fermi-Dirac-Amaldi model potential with 
optimized scaling parameters. These CI expansions included the 
configuration $\mathrm{1s^2\{2s^2, 2s2p, 2p^2\}}$
and all $\mathrm{1s^2\{2s,2p\}}nl$ configurations with \hbox{$n=3-7$} 
and \hbox{$l=0-5$} for \hbox{$n=3-5$} 
and \hbox{$l=0-2$} for \hbox{$n=6,7$}.
The DARC calculation~\cite{aggarwal2016} used orbitals obtained with the
GRASP (General-purpose Relativistic Atomic Structure Package) 
program~\cite{dyal1989,parpia1996} in the ``extended average level'' approximation.
Those CI expansions included the same set of configurations, albeit in
the $jj$-coupling scheme.

Note that Eq.~(\ref{eq:Phi}) is not just a minimal CI expansion, but it is  
enlarged by the inclusion of 
all $\mathrm{2s}nl$, $\mathrm{2p}nl$, $\mathrm{3s}nl$, $\mathrm{3p}nl$ 
and $\mathrm{3d}nl$ configurations, respectively, 
where $nl$ are the physical bound orbitals as well as the pseudo-orbitals 
that are used to discretize the continuum.
Consequently, we include all important single and double promotions for 
the $\mathrm{2s}nl$ and $\mathrm{2p}nl$ 
states under consideration.  This is very important for an accurate description of the 
valence correlation between the two outer electrons above the $\mathrm{1s^2}$ core. 
Furthermore, we directly include the term dependence of the one-electron wave functions. 
This term dependence is well known to be important for neutral Be, and it may be  
significant for other \hbox{Be-like} ions as well.

The expansion (\ref{eq:Phi}) generates an extensive spectrum of the $\mathrm{N}^{3+}$ 
ionic target, including the continuum.
This spectrum was reduced in the present calculations by limiting the maximum total
electronic angular momentum to $J=6$.
Altogether, we obtained $1400$ bound and continuum levels of $\mathrm{N}^{3+}$.
From these 
we selected for the close-coupling expansion just
the same 238 levels as in the previous works \cite{fernandez-menchero2014a,aggarwal2016}.

In Table~\ref{tab:energies} we list the excitation energies calculated with
the BSR method for the lowest 40~levels.
We compare them with the AS and GRASP 
\hbox{results~\cite{fernandez-menchero2014a,aggarwal2016}}, 
as well as the recommended values tabulated in the NIST database~\cite{moore1993}.
(The complete table of states can be found in the supplemental online material.)
As seen from the tables, our excitation energies agree much closer with experiment 
than either the AS or GRASP energies, especially for the low-lying $n=2$ states.
This is due to the more extensive CI expansions in our approach.
As noted in~\cite{aggarwal2016}, there is room for improving the accuracy of 
the energy levels by including pseudo-orbitals in the target wave functions, 
but it may lead to pseudo-resonance structure in the final collision strengths~\cite{burke2011}. 
Therefore, both Aggarwal {\em et al.}~\cite{aggarwal2016} and 
Fernandez-Menchero {\em et al.}~\cite{fernandez-menchero2014a} 
avoided this approach. 
The present BSR calculation, due to the use of non-orthogonal orbitals for
both the bound and continuum spectrum, is free from these restrictions.

Table \ref{tab:energiestop} shows the upper part of the spectrum to illustrate
how the states included are embedded into the full spectrum.
Above level \#87 ($\mathrm{2s\,6d\,^3D_3}$), there are levels (e.g., $\mathrm{2s\,6 \{f,g,h\}}$)
that are not found in the CI or CC expansions 
of~\cite{fernandez-menchero2014a,aggarwal2016}. 
Since the close-lying states may be strongly mixed, the expansions 
of~\cite{fernandez-menchero2014a} and~\cite{aggarwal2016} neglect possibly
important contributions to the wave functions in all levels above \#87.
In addition, the $\mathrm{2p}nl$ states, starting at the 
multiplet $\mathrm{2p5s\,^3P^{\rm o}}$, are located above the ionization threshold 
and hence are expected to strongly interact with the $\mathrm{2s}kl$ continuum. 
Our expansions~(\ref{eq:Phi}) include these interactions through the
continuum pseudostates, whereas they are completely omitted in 
the CI expansions used in~\cite{fernandez-menchero2014a,aggarwal2016}.

\begin{table*}
\caption{Selected high-lying energy levels of $\mathrm{N}^{3+}$.}
\label{tab:energiestop}
\begin{indented}
\item[\hspace{0.1\textwidth}]\begin{tabular}{rllrr@{\,}r@{\quad}r@{\,}r@{\quad}r@{\,}r@{}}
\br
 $i$   &              Conf.  &                Level &        NIST &        BSR & ($  \%$) &      GRASP & ($  \%$) &         AS & (  $\%$) \\
\mr
 $ 85$ & $\mathrm{2s\,6d}$  & $\mathrm{^3D_1^{e}}$ & $575030.5$  & $573968.2$ & ($-0.2$) & $573321.5$ & ($-0.3$) & $571523.0$ & ($-0.6$) \\
 $ 86$ & $\mathrm{2s\,6d}$  & $\mathrm{^3D_2^{e}}$ & $575030.5$  & $573969.0$ & ($-0.2$) & $573321.5$ & ($-0.3$) & $571524.0$ & ($-0.6$) \\
 $ 87$ & $\mathrm{2s\,6d}$  & $\mathrm{^3D_3^{e}}$ & $575030.5$  & $573970.3$ & ($-0.2$) & $573322.6$ & ($-0.3$) & $571525.0$ & ($-0.6$) \\
 $ 88$ & $\mathrm{2s\,6g}$  & $\mathrm{^3G_3^{e}}$ & $575809.4$  & $574725.8$ & ($-0.2$) & $      - $ &  $    $  & $      - $ &  $    $  \\
 $ 89$ & $\mathrm{2s\,6g}$  & $\mathrm{^3G_4^{e}}$ & $575809.4$  & $574726.4$ & ($-0.2$) & $      - $ &  $    $  & $      - $ &  $    $  \\
 $ 90$ & $\mathrm{2s\,6g}$  & $\mathrm{^1G_4^{e}}$ & $575807.9$  & $574728.2$ & ($-0.2$) & $      - $ &  $    $  & $      - $ &  $    $  \\
 $ 91$ & $\mathrm{2s\,6g}$  & $\mathrm{^3G_5^{e}}$ & $575809.4$  & $574728.2$ & ($-0.2$) & $      - $ &  $    $  & $      - $ &  $    $  \\
 $ 92$ & $\mathrm{2s\,6f}$  & $\mathrm{^3F_2^{o}}$ & $      - $  & $574842.5$ &  $    $  & $      - $ &  $    $  & $      - $ &  $    $  \\
 $ 93$ & $\mathrm{2s\,6f}$  & $\mathrm{^3F_3^{o}}$ & $      - $  & $574843.4$ &  $    $  & $      - $ &  $    $  & $      - $ &  $    $  \\
 $ 94$ & $\mathrm{2s\,6f}$  & $\mathrm{^3F_4^{o}}$ & $      - $  & $574844.7$ &  $    $  & $      - $ &  $    $  & $      - $ &  $    $  \\
 $ 95$ & $\mathrm{2s\,6d}$  & $\mathrm{^1D_2^{e}}$ & $575872.4$  & $574867.5$ & ($-0.2$) & $574527.5$ & ($-0.2$) & $572601.0$ & ($-0.6$) \\
 $ 96$ & $\mathrm{2s\,6f}$  & $\mathrm{^1F_3^{o}}$ & $575999.3$  & $574924.7$ & ($-0.2$) & $      - $ &  $    $  & $      - $ &  $    $  \\
 $ 97$ & $\mathrm{2s\,6h}$  & $\mathrm{^3H_4^{o}}$ & $576042.9$  & $574956.5$ & ($-0.2$) & $      - $ &  $    $  & $      - $ &  $    $  \\
 $ 98$ & $\mathrm{2s\,6h}$  & $\mathrm{^3H_5^{o}}$ & $576042.9$  & $574956.6$ & ($-0.2$) & $      - $ &  $    $  & $      - $ &  $    $  \\
 $ 99$ & $\mathrm{2s\,6h}$  & $\mathrm{^1H_5^{o}}$ & $576042.9$  & $574956.8$ & ($-0.2$) & $      - $ &  $    $  & $      - $ &  $    $  \\
 $100$ & $\mathrm{2s\,6h}$  & $\mathrm{^3H_6^{o}}$ & $576042.9$  & $574956.8$ & ($-0.2$) & $      - $ &  $    $  & $      - $ &  $    $  \\
 $101$ & $\mathrm{2p\,4s}$  & $\mathrm{^3P_0^{o}}$ & $577957.8$  & $577512.3$ & ($-0.1$) & $577188.6$ & ($-0.1$) & $575628.0$ & ($-0.4$) \\
 $102$ & $\mathrm{2p\,4s}$  & $\mathrm{^3P_1^{o}}$ & $578045.4$  & $577589.6$ & ($-0.1$) & $577267.7$ & ($-0.1$) & $575712.0$ & ($-0.4$) \\
\mr    
 $266$ & $\mathrm{2s\,11d}$ & $\mathrm{^3D_3^{e}}$ & $610166.0$  & $620383.2$ & ($ 1.7$) & $      - $ &  $    $  & $      - $ &  $    $  \\
 $267$ & $\mathrm{2s\,11d}$ & $\mathrm{^1D_2^{e}}$ & $610287.2$  & $620914.7$ & ($ 1.7$) & $      - $ &  $    $  & $      - $ &  $    $  \\
 $268$ & $\mathrm{2s\,12k}$ & $\mathrm{^3K_6^{o}}$ & $      - $  & $622346.8$ &  $    $  & $      - $ &  $    $  & $      - $ &  $    $  \\
\mr    
 \multicolumn{3}{l}{Ionization limit}              & $624866.0$  &            &          &            &          &            &          \\
\mr    
 $269$ & $\mathrm{2s\,ki}$  & $\mathrm{^3I_5^{e}}$ & $      - $  & $624777.0$ &  $    $  & $      - $ &  $    $  & $      - $ &  $    $  \\
 $270$ & $\mathrm{2s\,ki}$  & $\mathrm{^3I_6^{e}}$ & $      - $  & $624777.0$ &  $    $  & $      - $ &  $    $  & $      - $ &  $    $  \\
 $271$ & $\mathrm{2s\,ki}$  & $\mathrm{^1I_6^{e}}$ & $      - $  & $624777.1$ &  $    $  & $      - $ &  $    $  & $      - $ &  $    $  \\
 $272$ & $\mathrm{2p\,5s}$  & $\mathrm{^3P_0^{o}}$ & $626355.0$  & $625844.3$ & ($-0.1$) & $625435.8$ & ($-0.1$) & $623873.0$ & ($-0.4$) \\
 $273$ & $\mathrm{2p\,5s}$  & $\mathrm{^3P_1^{o}}$ & $626445.3$  & $625915.4$ & ($-0.1$) & $625511.5$ & ($-0.1$) & $623952.0$ & ($-0.4$) \\
 $274$ & $\mathrm{2p\,5s}$  & $\mathrm{^3P_2^{o}}$ & $626611.9$  & $626089.6$ & ($-0.1$) & $625684.9$ & ($-0.1$) & $624133.0$ & ($-0.4$) \\
 $275$ & $\mathrm{2p\,5s}$  & $\mathrm{^1P_1^{o}}$ & $628546.9$  & $627025.1$ & ($-0.2$) & $627423.1$ & ($-0.2$) & $625792.0$ & ($-0.4$) \\
\br
\end{tabular}

\item[]Key: $i$, level index; Conf., Level: Configuration and multiplet, largest weight;
BSR: calculated excitation energy relative to the ground state, present work;
NIST: recommended value from the NIST data base \cite{moore1993};
AS~\cite{fernandez-menchero2014a}.
GRASP~\cite{aggarwal2016};
$\%$: relative deviation from the recommended values.
All energies are given in $\mathrm{cm}^{-1}$.

\end{indented}
\end{table*}

A standard assessment regarding the quality of structure calculations involves a
comparison of the oscillator strengths~$f$, 
or the closely related~$gf$, which is the value of~$f$ multiplied by the statistical
weight ($2J+1$) of the initial level.
The $gf$-values are the same for excitation and de-excitation.
Note that oscillator strengths themselves are also very important
for plasma modeling.

\begin{figure}
   \includegraphics[width=1.00\columnwidth,clip]{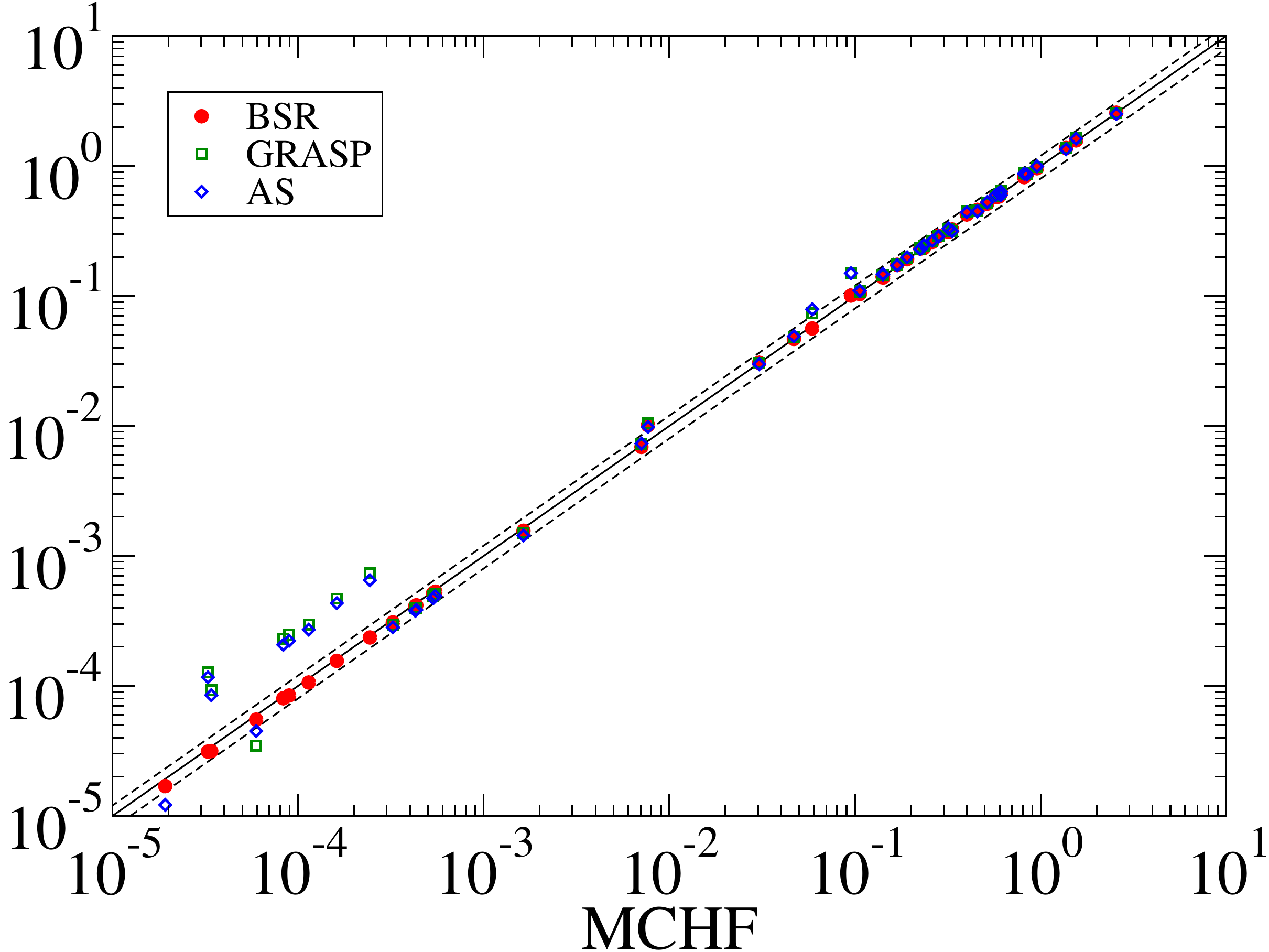}
   \caption{\label{fig:MCHF} Comparison of $gf$ from the BSR, 
   DARC, and AS calculations with the MCHF results~\cite{froese-fischer2004}.
   $x$-axis: $gf$ results for a certain transition calculated with the MCHF
   method \cite{froese-fischer2004};
   $y$-axis: $gf$ results for the same transition calculated with 
   BSR (present work), GRASP~\cite{aggarwal2016},
   and AS~\cite{fernandez-menchero2014a}.
   The dashed lines show the bands corresponding to a $20\%$ deviation from the MCHF results.}
\end{figure}

We begin by comparing results for the transitions between the lowest 20 states, where 
there are numerous data from other calculations.
In particular, Froese-Fischer and coworkers~\cite{froese-fischer2004} presented 
extensive MCHF calculations with a large set 
of configurations and a careful analysis of convergence.
It is generally accepted that this work represents the most accurate calculation 
for the structure of the lowest excited levels of several Be-like ions, including $\mathrm{N}^{3+}$.
Figure~\ref{fig:MCHF} shows a comparison of the MCHF oscillator strengths
with the results from the BSR, DARC, and AS models for
transitions between the lowest 20 levels.
We see very good agreement between the BSR and MCHF results for all transitions, 
including the very weak ones with small \hbox{$gf$-values}. 
This suggests similarly accurate configuration mixing in both calculations. 
The DARC and AS results only show good agreement with MCHF for strong transitions
with $gf$-values larger than $10^{-3}$, whereas they differ considerably from MCHF and BSR for weak 
transitions, where configuration mixing plays an important role. 
This mixing primarily depends on correlation corrections, 
included in the MCHF and BSR calculations, rather than on the spin-orbit mixing that is accounted for
in all models.

Only a few experimental data for the 
oscillator strengths
or transition rates
in~$\mathrm{N}^{3+}$ are available to 
assess the accuracy of the different predictions independently of the MCHF results.
Table~\ref{tab:Acompexp} shows a comparison with
experiment for the two lines from the $\mathrm{2s2p}$ configuration.
For the strong resonant $\mathrm{2s^2\,^1S_0 - 2s2p\,^1P_1^{o}}$ transition
all calculations agree closely with each other and with experiment.
For the weak intercombination transition $\mathrm{2s^2\,^1S_0-2s2p\,^3P_1^{o}}$,
the agreement is not close, and the presumably best model (MCHF) produces a 
value that deviates from experiment by more than twice the estimated experimental uncertainty.
Note, however, that the relative experimental uncertainty is large (more than 25\%).  Hence,
while the GRASP number appears to agree best with the experimental result, the BSR prediction deviates 
from the latter by just about the published uncertainty and hence is essentially consistent with experiment
as well.  This clearly shows that theoretical results for weak transitions are very sensitive to the model
employed, due to the increasing likelihood of cancellations from contributing matrix elements.
Experimental data for such transitions are generally uncertain as well.
Based on long-term experience in extensive structure calculations, we continue to believe 
that the MCHF result is the most reliable, due to the size of the expansion and the 
systematic convergence checks performed.  Unfortunately, only an independent experiment and/or
even more extensive structure calculations, using as much as possible similar semi-relativistic and 
full-relativistic expansions, would be able to shed more light on this problem.

\begin{table}
\caption{Comparison of oscillator strengths and transition rates.}
\label{tab:Acompexp}
\begin{indented}
\item[\hspace{10.0truemm}]
\begin{tabular}{@{}ll}
   \br
   $\mathrm{2s^2\,^1S_0 - 2s2p\,^1P_1^{o}}$ & $f$-factor \\
   \mr
   Present work                      & $0.6268$ \\
   Expt. \cite{engstrom1981}          & $0.620 \pm 0.022$  \\
   MCHF \cite{froese-fischer2004}    & $0.610$  \\
   GRASP \cite{aggarwal2016}         & $0.6391$ \\
   AS \cite{fernandez-menchero2014a} & $0.6326$ \\
   \br
\end{tabular}
\item[\hspace{10.0truemm}]
\begin{tabular}{@{}ll}
   \br
   $\mathrm{2s^2\,^1S_0 - 2s2p\,^3P_1^{o}}$ & Transition rate $A$ (s$^{-1}$). \\ 
   \mr
   Present work                      & $4.772 \times 10^{-7}$ \\
   Expt.~\cite{doerfert1996}          & $(3.8 \pm 0.9) \times 10^{-7}$ \\
   MCHF~\cite{froese-fischer2004}    & $5.755 \times 10^{-7}$ \\
   GRASP~\cite{aggarwal2016}         & $4.030 \times 10^{-7}$ \\
   AS~\cite{fernandez-menchero2014a} & $5.781 \times 10^{-7}$ \\
   \br
\end{tabular}
\end{indented}
\end{table}

\begin{figure}
\centering
      \includegraphics[width=1.00\columnwidth,clip]{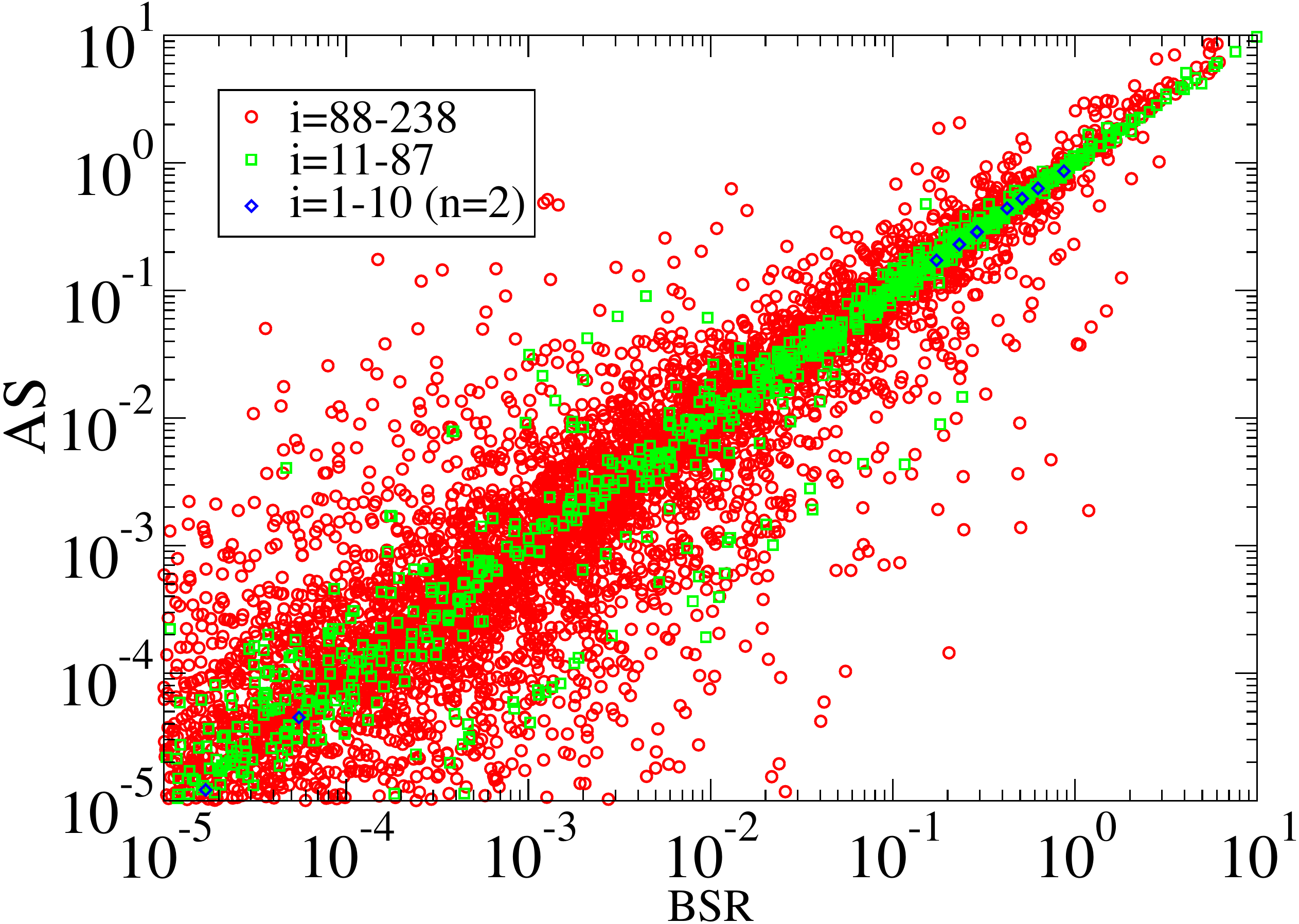}
      \includegraphics[width=1.00\columnwidth,clip]{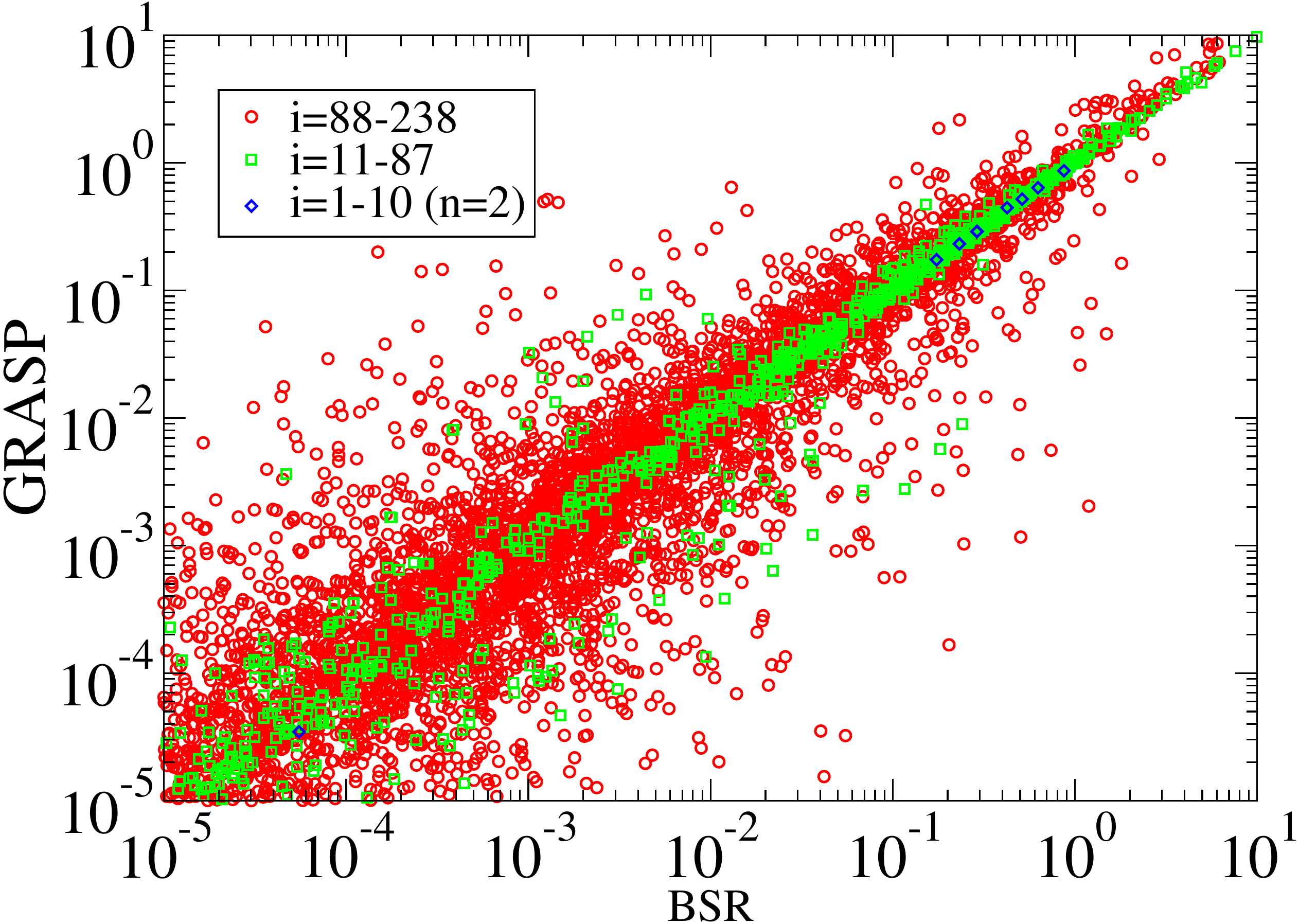}
   \caption{Comparison of $gf$-values for the $\mathrm{N}^{3+}$ structure obtained in the
      present work with~\cite{fernandez-menchero2014a} (top panel)
      and~\cite{aggarwal2016} (bottom panel).
      $x$-axis: $gf$ results for a certain transition calculated with the BSR
      method, present work;
      \hbox{$y$-axis}: $gf$ results for the same transition calculated with 
      AS (upper panel \cite{fernandez-menchero2014a}) and 
      GRASP (lower panel \cite{aggarwal2016}).
      $\diamond$: transitions within the $n=2$ electronic shell, upper level up to 10;
      $\square$: transitions with upper level between 11 and 87;
      $\circ$: transitions with upper level above 88.}
   \label{fig:gfn3}
\end{figure}

For further comparison of the present target structure with previous works,
we show in Fig.~\ref{fig:gfn3} the \hbox{$gf$-values} of BSR vs.\ DARC and ICFT
for the electric dipole-allowed transitions between all 238 IC levels
included in the scattering calculations.
(A complete list of the radiative parameters from the present BSR calculations 
is provided in the online supplemental material.)
The differences in the \hbox{$gf$-values} are relatively small, on average less than~$50\%$,
for most transitions with an upper level up to~\#87.
There are a few exceptions for weak transitions, where
the differences can be as large as several orders of magnitude.
In these cases, the electric dipole transition takes place through spin-orbit
mixing between configurations of small weight in the CI expansion.
A slight change in the mixing coefficients can then change
the \hbox{$gf$-value} considerably.
In practice, an accurate description of such weak intercombination transitions
can only be achieved in specially designed calculations,
with careful consideration of each individual transition.
However, since such transitions usually have very small $gf$-values,
they are unlikely to have a significant effect in plasma modeling.

As seen from Fig.~\ref{fig:gfn3}, the differences in the $gf$-values
obtained by the various methods considerably increase for levels above~\#87.
This applies to both the BSR/DARC and BSR/AS comparisons, with approximately
the same level of agreement or disagreement.
Note that the BSR expansions for these levels
are very different from those used in the DARC and AS calculations.
They are more extensive and include configuration
mixing with the additional 
$\mathrm{\{3s,3p,3d\}}nl$ bound states, as well as
the interaction with the continuum (see Table~\ref{tab:energiestop}).

\begin{figure}
\centering
   \includegraphics[width=1.00\columnwidth,clip]{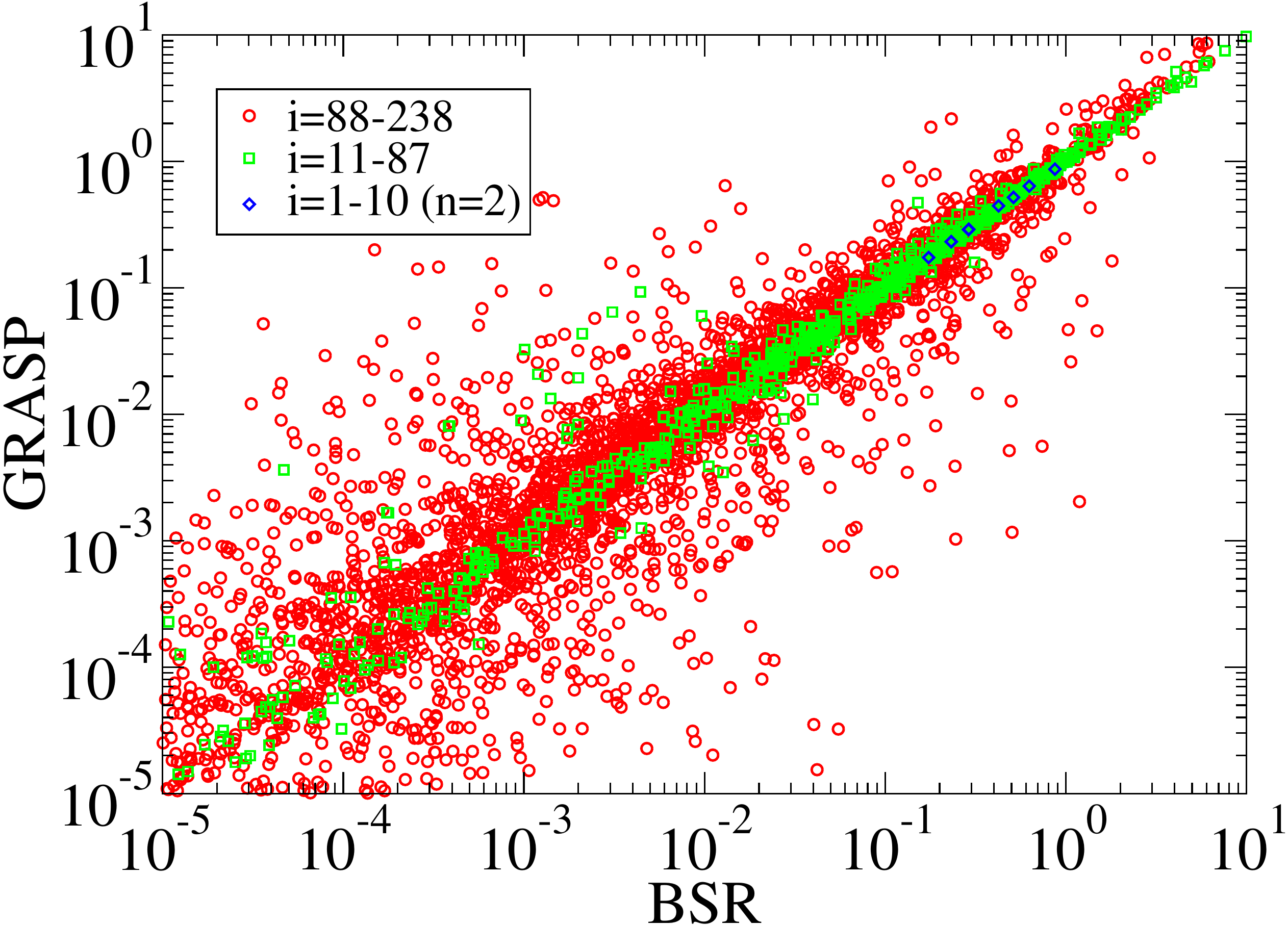}
   \caption{As in Fig.~\ref{fig:gfn3}, but limited to spin-conserving transitions.
      }
   \label{fig:gfn3_nsx}
\end{figure}

As a result, the spin-orbit mixing is also described very differently for
these states.
These differences in the \hbox{$gf$-values} are expected to
cause deviations of the same order in the final results
for the effective collisions strength~$\Upsilon$, especially at high temperatures.
To illustrate the effect of the different relativistic approaches on the
predicted target structure, we show in Fig.~\ref{fig:gfn3_nsx}
the same \hbox{$gf$-values}, but this time limited to spin-conserving
transitions.
Relativistic effects, specifically the spin-orbit 
interaction, should be negligible in these cases.
Since the dispersion seen in the figure for spin-conserving transitions
is similar to that for the spin-changing ones,
the way to include relativistic 
terms in the hamiltonian affects to a much lesser extent the predicted target
properties than the method used to expand the wave functions.

\begin{figure}
   \includegraphics[width=1.00\columnwidth,clip]{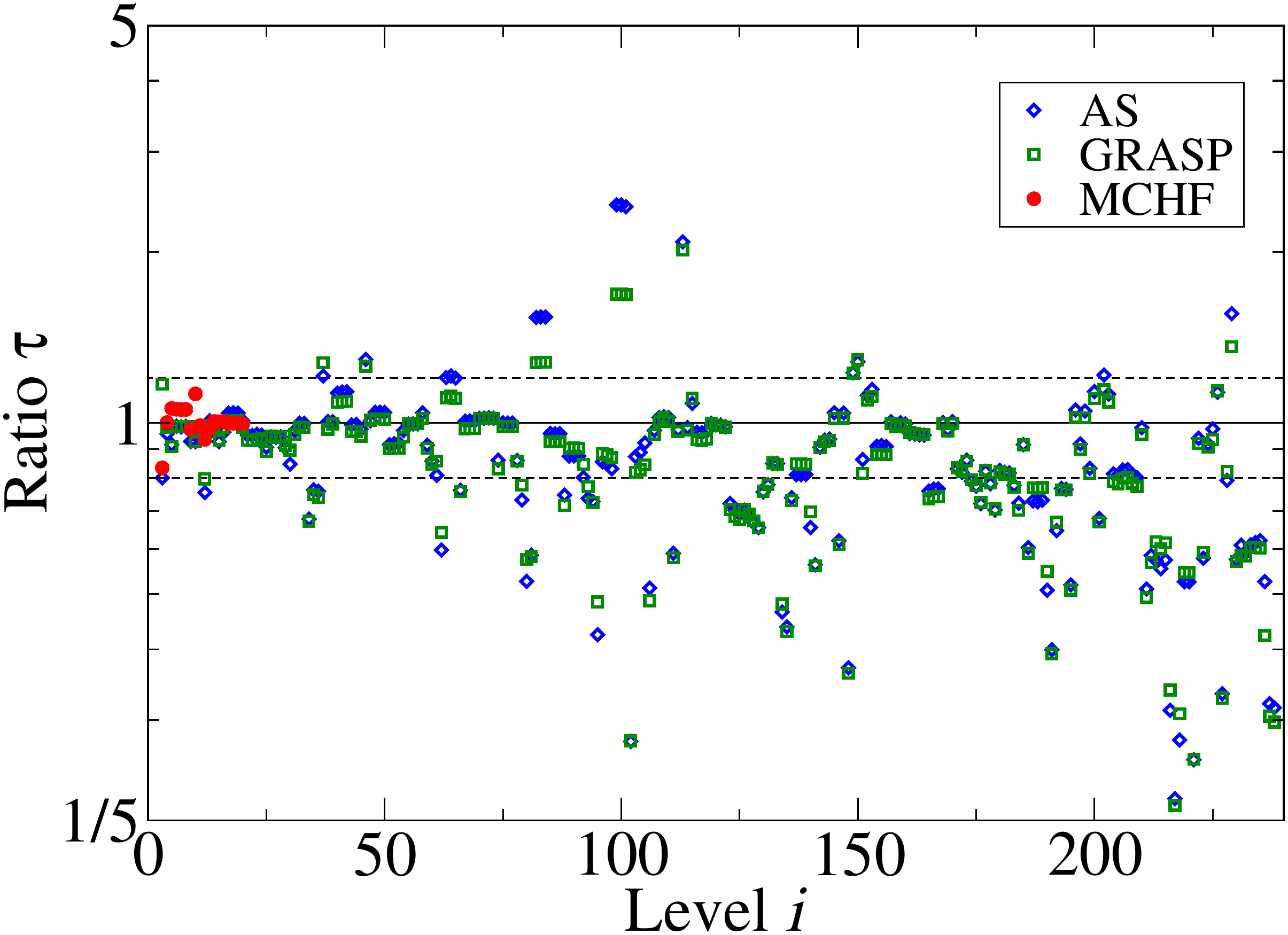}
   \caption{\label{fig:tau} Lifetimes from the MCHF, DARC, and 
            AS calculations normalized to the BSR predictions.
            The dashed lines indicate the band of $20\%$ deviation.
            $\diamond$: $\tau(\mathrm{AS})/\tau(\mathrm{BSR})$; 
            $\square$: $\tau(\mathrm{GRASP})/\tau(\mathrm{BSR})$; 
            $\bullet$: $\tau(\mathrm{MCHF})/\tau(\mathrm{BSR})$.
            }
\end{figure}

Variations in the $f$-values also results in different
predictions of the lifetimes.  
A comparison of lifetimes from the various calculations is shown in Fig.~\ref{fig:tau}. 
The ratios of the MCHF, DARC, and AS results to the BSR predictions are close to~$1$ for the low-lying 
states, but they differ considerably for some states in the upper-part of the spectrum. 
This indicates, once more, the sensitivity of the results to the different configuration expansions  
used in the individual methods.
Since it is unlikely that the configuration expansions are near  
convergence in either the DARC or the AS calculations, the accuracy of the 
predicted collision strengths from these models is questionable.  This will be further discussed below.

\section{Collision Calculations}
\label{sec:scattering}

For the scattering calculations we used the \hbox{B-spline} \hbox{R-matrix} 
code of Zatsarinny~\cite{zatsarinny2006}. 
The distinctive feature of the method is the use of \hbox{B-splines} as a universal 
basis to represent the scattering orbitals in the inner region of $r \le a$.
The principal advantage of \hbox{B-splines} is that they form an effectively complete 
basis, and hence no Buttle correction 
\cite{burke2011,buttle1967} 
to the \hbox{R-matrix} is needed in this case. 
The amplitudes of the wave functions at the boundary, which are required for the
evaluation of the \hbox{R-matrix}, are given by the coefficient of the last spline, 
which is the only spline with nonzero value at the boundary.

The other important feature of the present code concerns the
orthogonality requirements for the one-electron radial functions.
We impose only limited orthogonality conditions on the continuum orbitals.
This avoids the $(N+1)$-electron bound-like configurations, which need to be included in 
standard \hbox{R-matrix} calculations to ensure the numerical completeness of the expansion~\cite{burke2011}.
It also allows us to use much more extensive multi-configuration expansions for 
the target states while avoiding the pseudo-resonance structure that may appear
in calculations with an extensive number of correlated orbitals.

The present close-coupling expansions include 238 target states,
the same ones as in~\cite{fernandez-menchero2014a,aggarwal2016}.
These states consist of all $LSJ$ levels arising
from the configurations $\mathrm{1s^2\,\{2s^2,2s\,2p,2p^2\}}$ and
$\mathrm{1s^2\,\{2s,2p\}}\,nl$ for $nl$ orbitals
3s, 3p, 3d, 4s, 4p, 4d, 4f, 5s, 5p, 5d, 5f, 5g, 6s, 6p, 6d, 7s, 7p, 7d.
Relativistic effects in the scattering calculations were incorporated
via the Breit-Pauli hamiltonian through the one-electron Darwin, mass correction,
and spin-orbit operators.
The computing resources needed in this approach are about an order of magnitude larger than in the ICFT formalism,
due to the size of the hamiltonian matrices that need to be diagonalized.
In~\cite{fernandez-menchero2014a} the CC calculation was carried out
in $LS$-coupling with only 130 terms, and afterwards the
\hbox{$\bm{K}$-matrixes} were recoupled to obtained the level-resolved collision strengths.

\begin{figure*}
\centering
      \includegraphics[width=0.33\textwidth,clip]{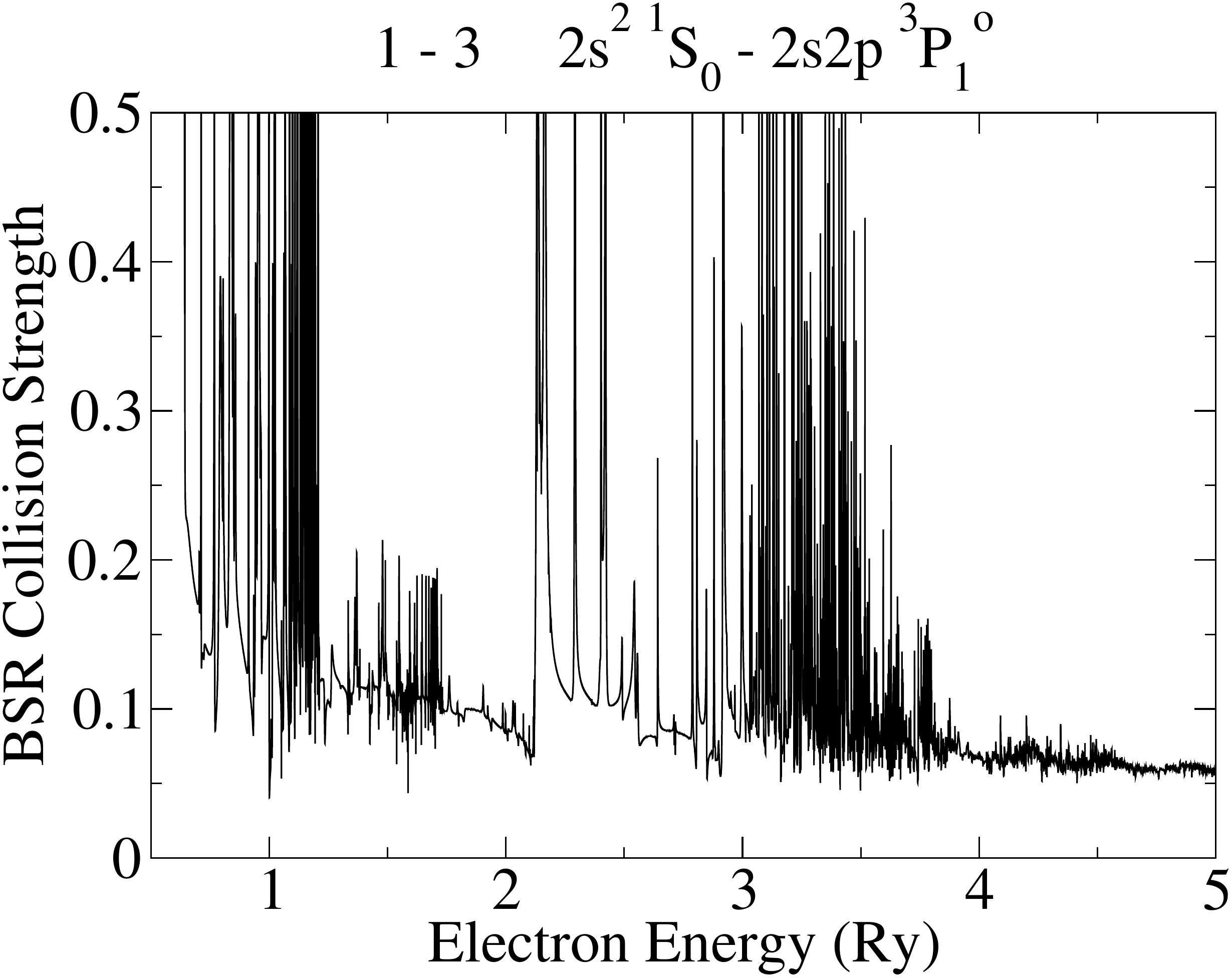}
      \includegraphics[width=0.33\textwidth,clip]{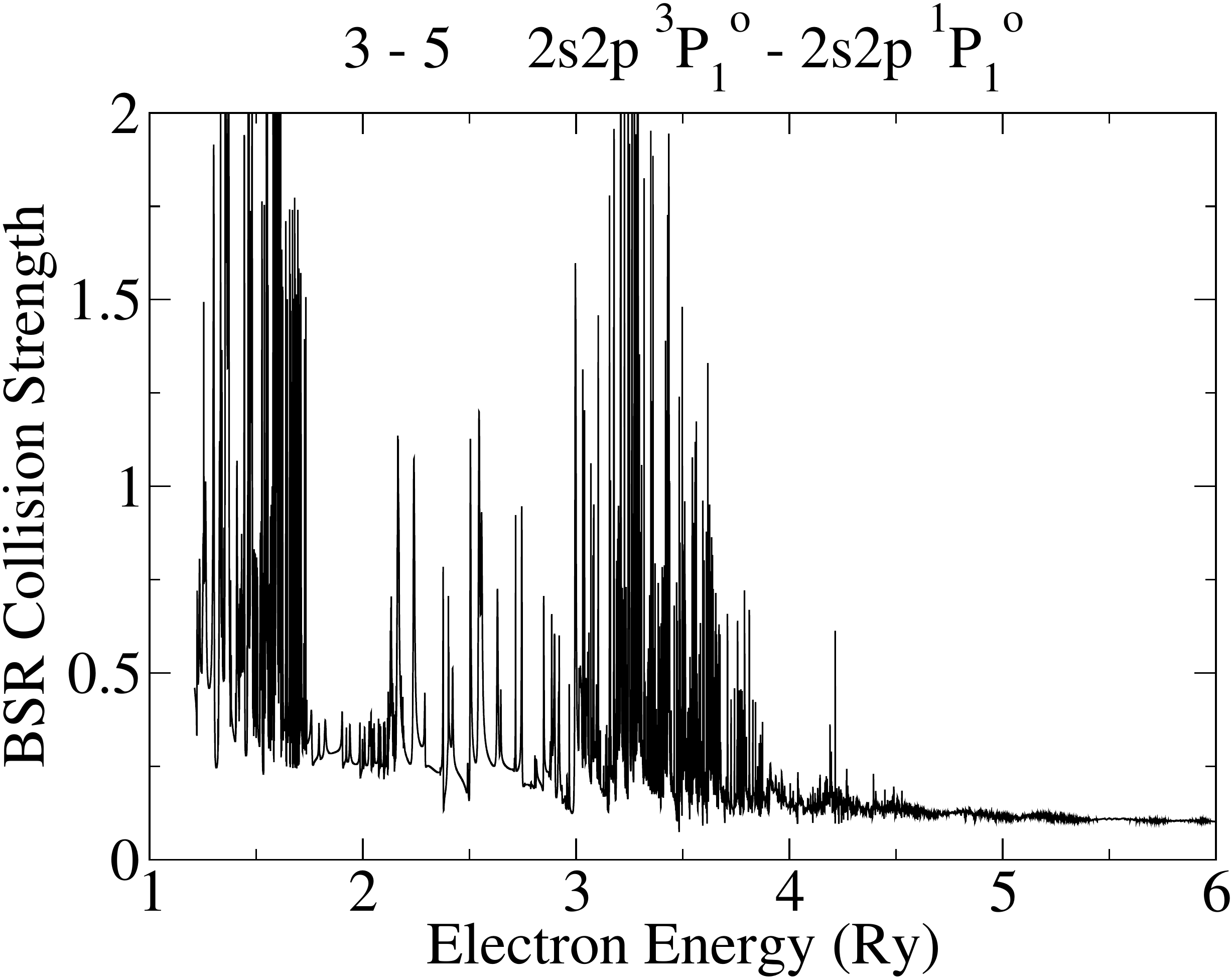}
      \includegraphics[width=0.33\textwidth,clip]{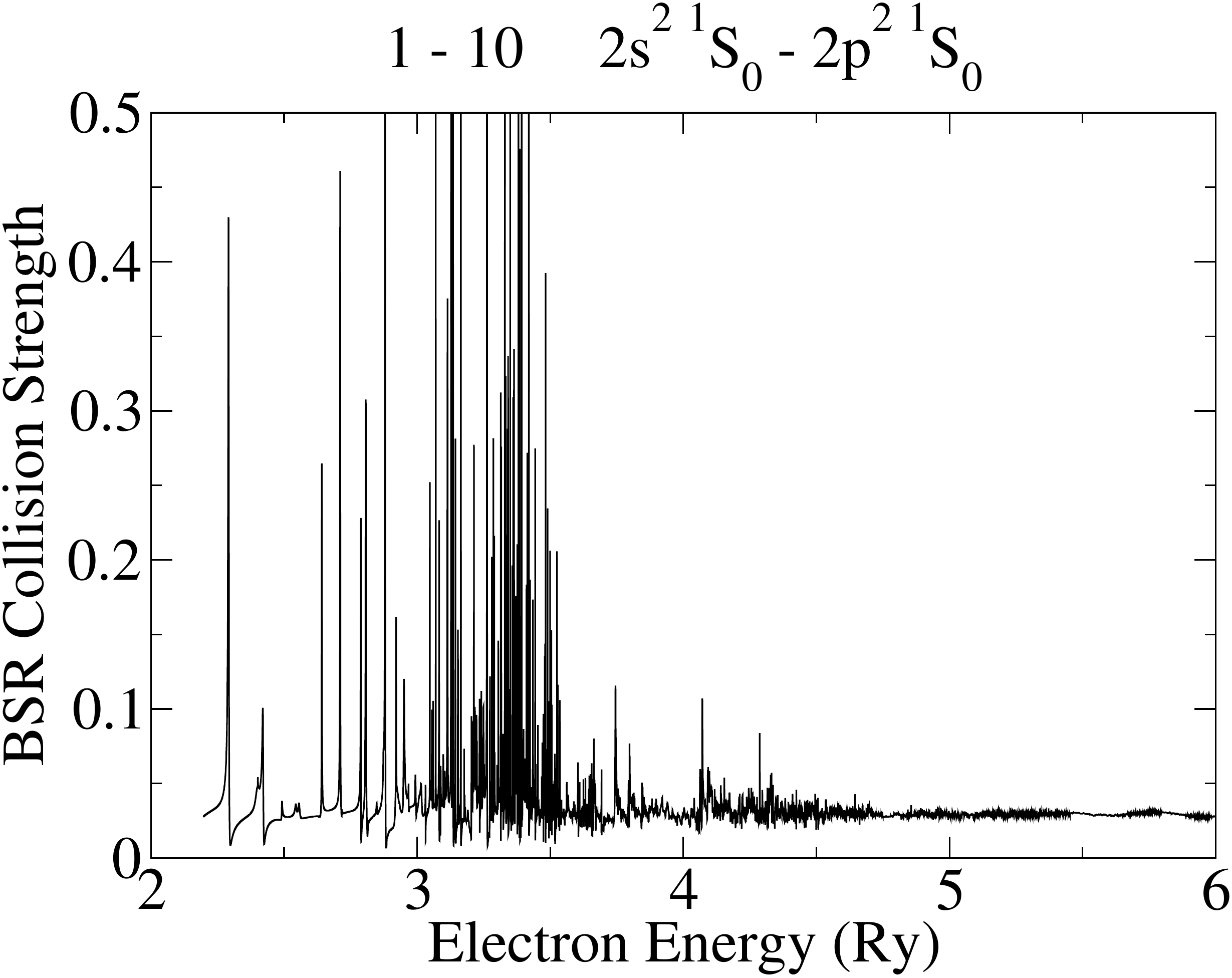}
      \includegraphics[width=0.33\textwidth,clip]{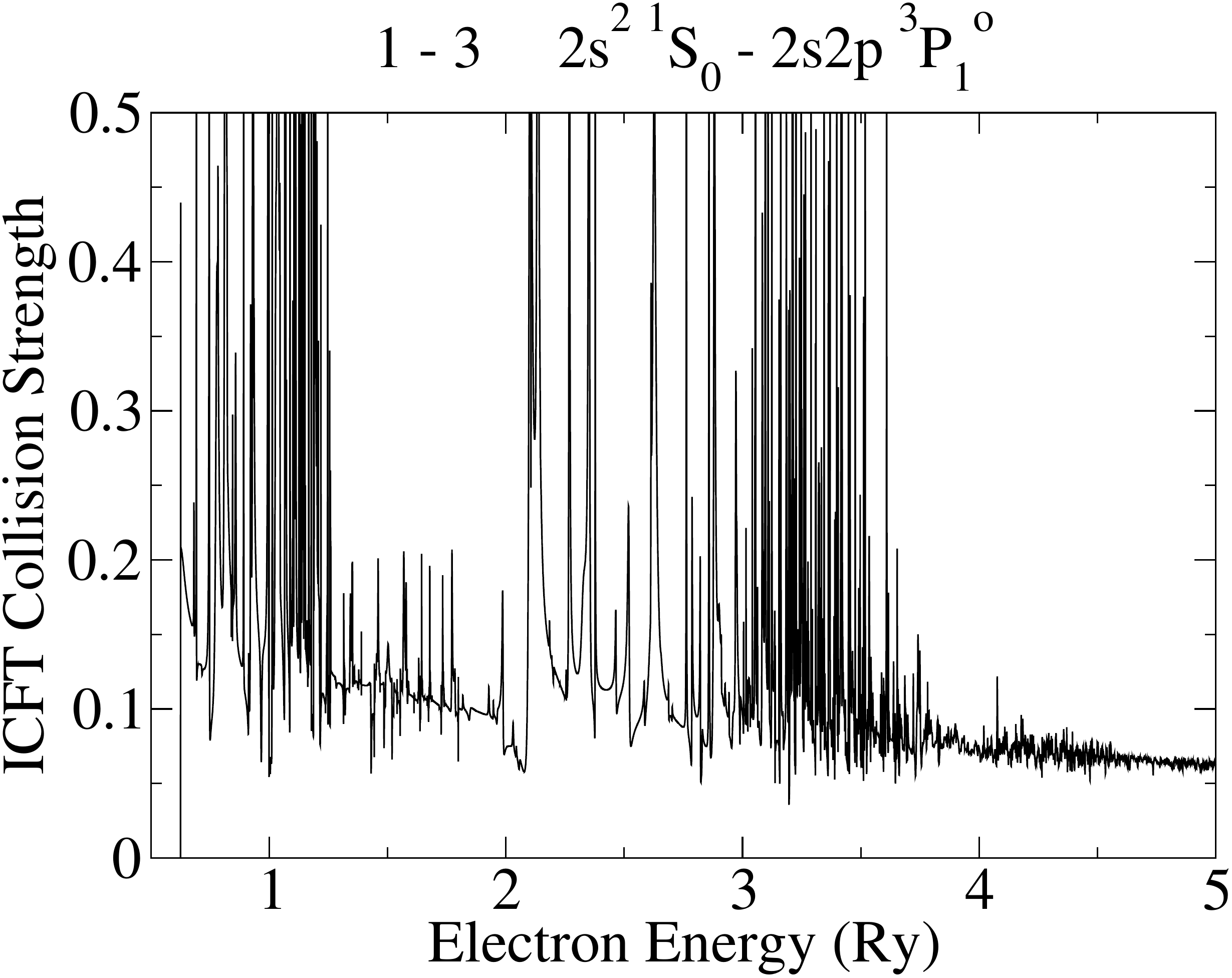}
      \includegraphics[width=0.33\textwidth,clip]{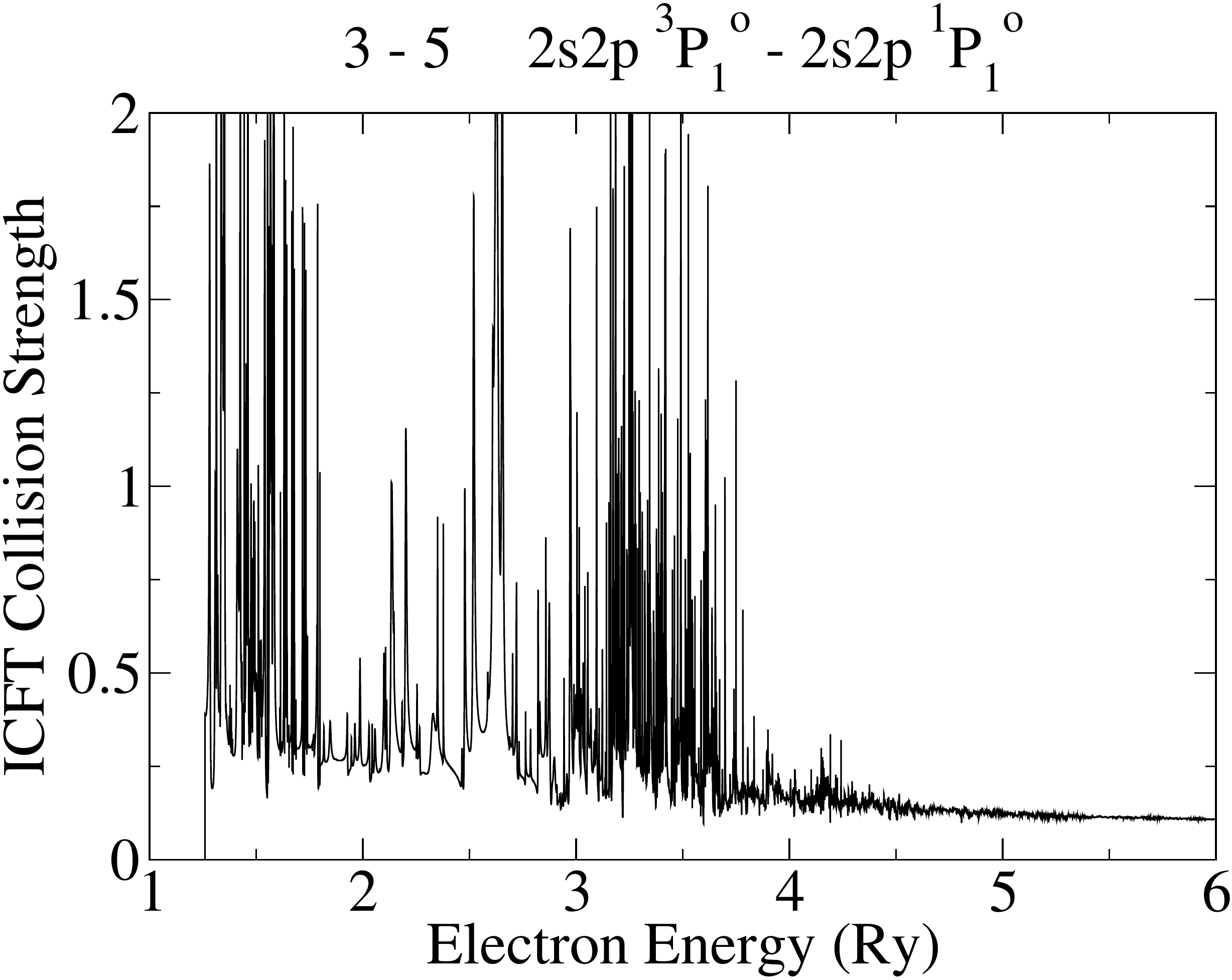}
      \includegraphics[width=0.33\textwidth,clip]{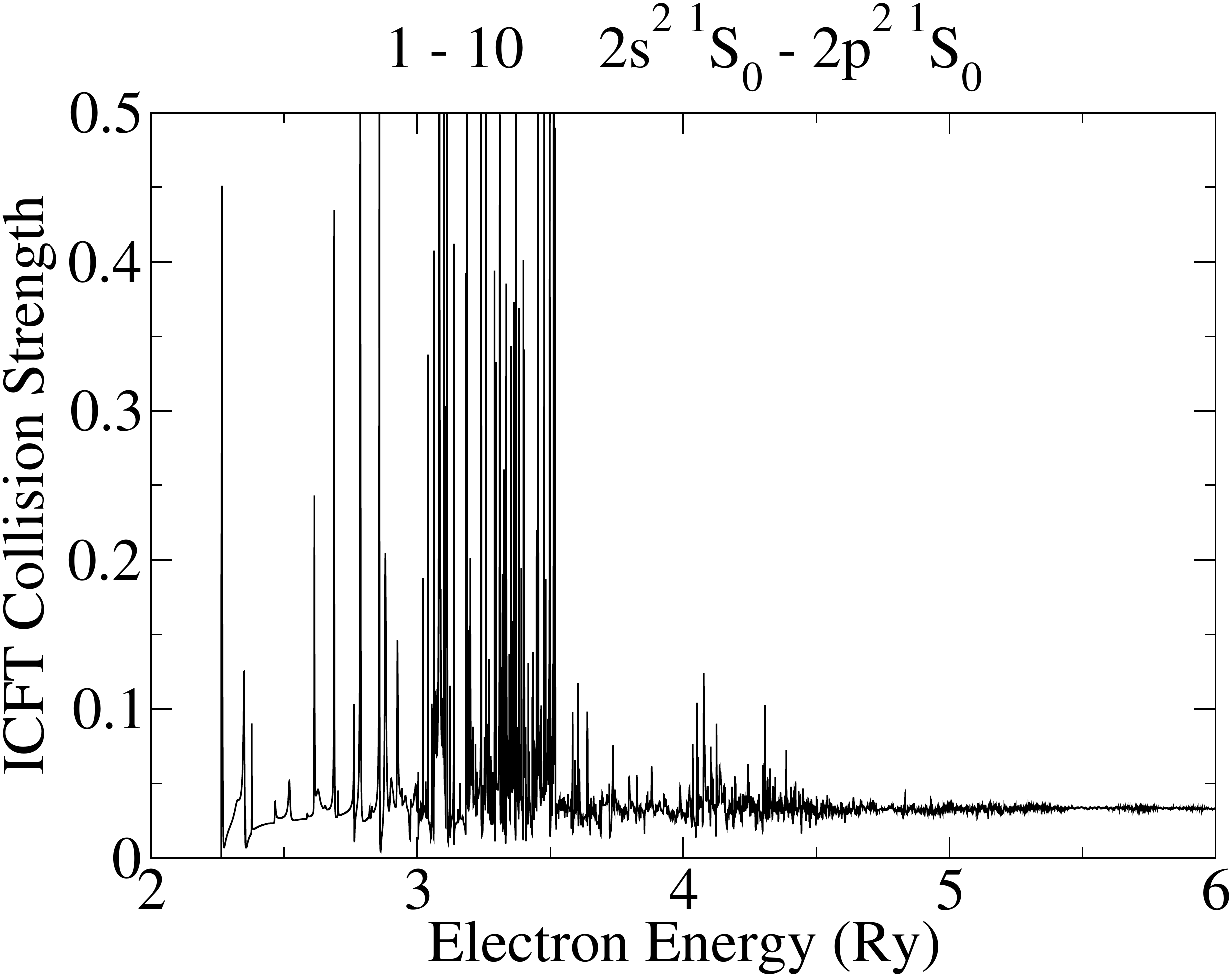}
   \caption{Collision strength $\Omega$ for electron-impact
      excitation of $\mathrm{N}^{3+}$ for selected transitions.
      Top panels: present BSR;
      bottom panels: ICFT~\cite{fernandez-menchero2014a}.}
   \label{fig:omega}
\end{figure*}

In the inner region, we used the same \hbox{B-spline} set as for the target 
CI structure described in Sect.~\ref{sec:structure}.
Numerical calculations were performed for 90 $LSJ$ partial waves up to 
$J_{\rm max} = 89/2$, for both even and odd parities.
The maximum number of channels in a partial wave was $1116$,
the maximum number of physical and correlated orbitals was $4837$,
and the maximum number of configurations was $83095$.
To account for the contributions from angular momenta higher than~$J_{\rm max} = 89/2$, we
used the Burgess sum rule~\cite{burgess1974} for optically allowed transitions
and a geometric series for all others~\cite{badnell2001a}.

For the outer region we employed a parallel version of the {\tt PSTGF}
program~\cite{berrington1987,badnell1999a}.
In the resonance regime for impact energies below the excitation energy of the highest level
included in the CC expansion, we used a fine energy step
of $10^{-5}\,z^2 \Ry$, with $z=3$ as the target ionic charge, to 
properly map those resonances.
For energies above the highest excitation threshold 
included in the CC expansion, the collision strengths vary
smoothly, and hence we chose a coarser step of $10^{-2}\,z^2 \Ry$.
Altogether, 61279 energies for the colliding electron were considered. 
We calculated collision strengths up to~$30 \Ry$, which is about five times the ionization
potential of $\mathrm{N}^{3+}$~\cite{moore1993}.
For even higher energies we extra\-polated using the well-known asymptotic energy dependence
for the various transitions.
To obtain effective collision strengths~$\Upsilon(T)$, we convoluted
$\Omega$ with a Maxwellian distribution for the electron
temperature~$T$, i.e., 
\begin{equation}
   \Upsilon_{i-j}(T)\ =\ \int_{E_{th}}^{\infty} \rd E\,\Omega_{i-j}(E)\,
   \exp \left( \frac{E-E_{th}}{kT} \right) \,.
\end{equation}
Here $E_{th}$ is the $i-j$ transition energy and $k$ is the Boltz\-mann constant.

\section{Results and Discussion}
\label{sec:results}

\begin{figure*} 
\centering
   \includegraphics[width=0.38\textwidth,clip]{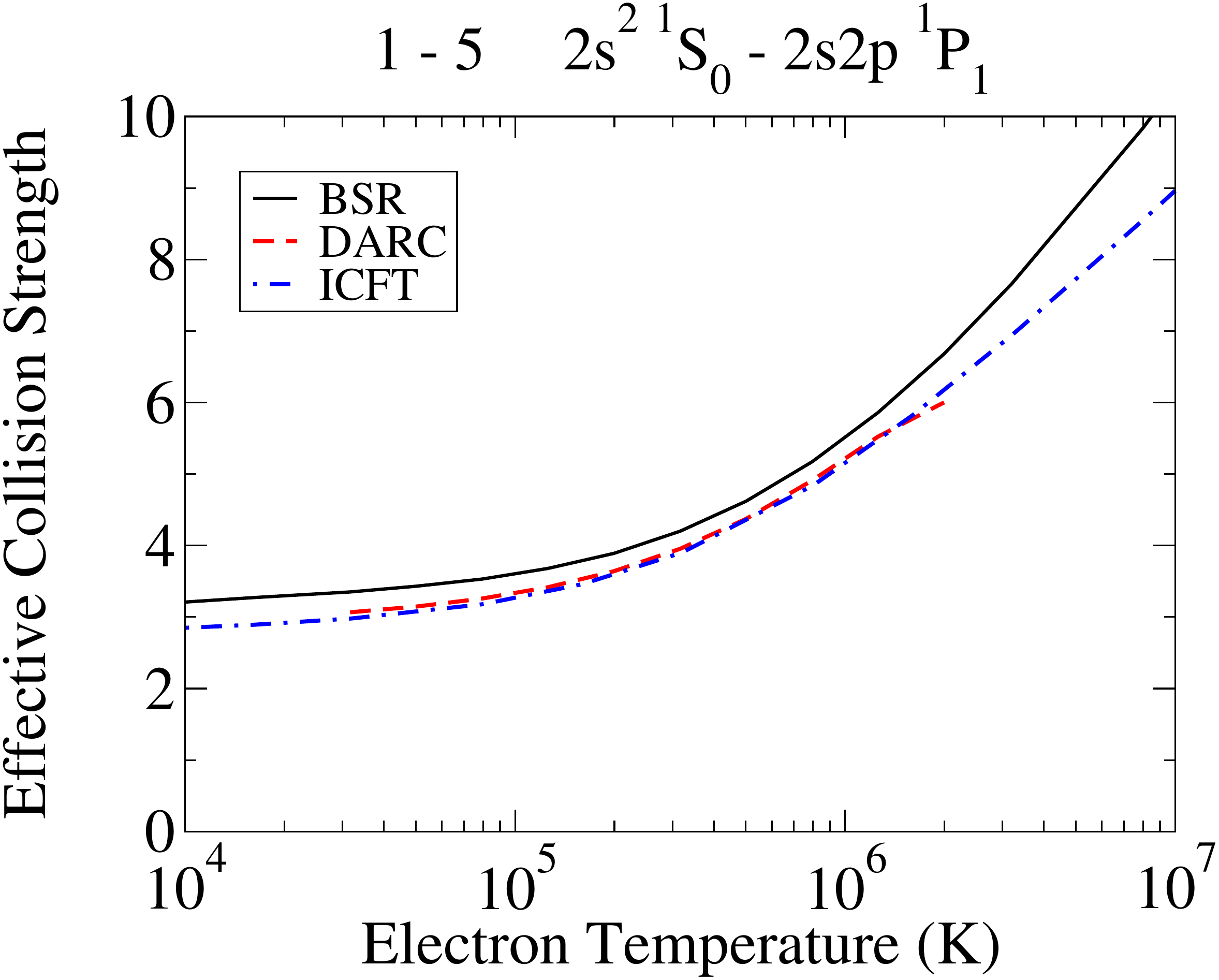}  \includegraphics[width=0.38\textwidth,clip]{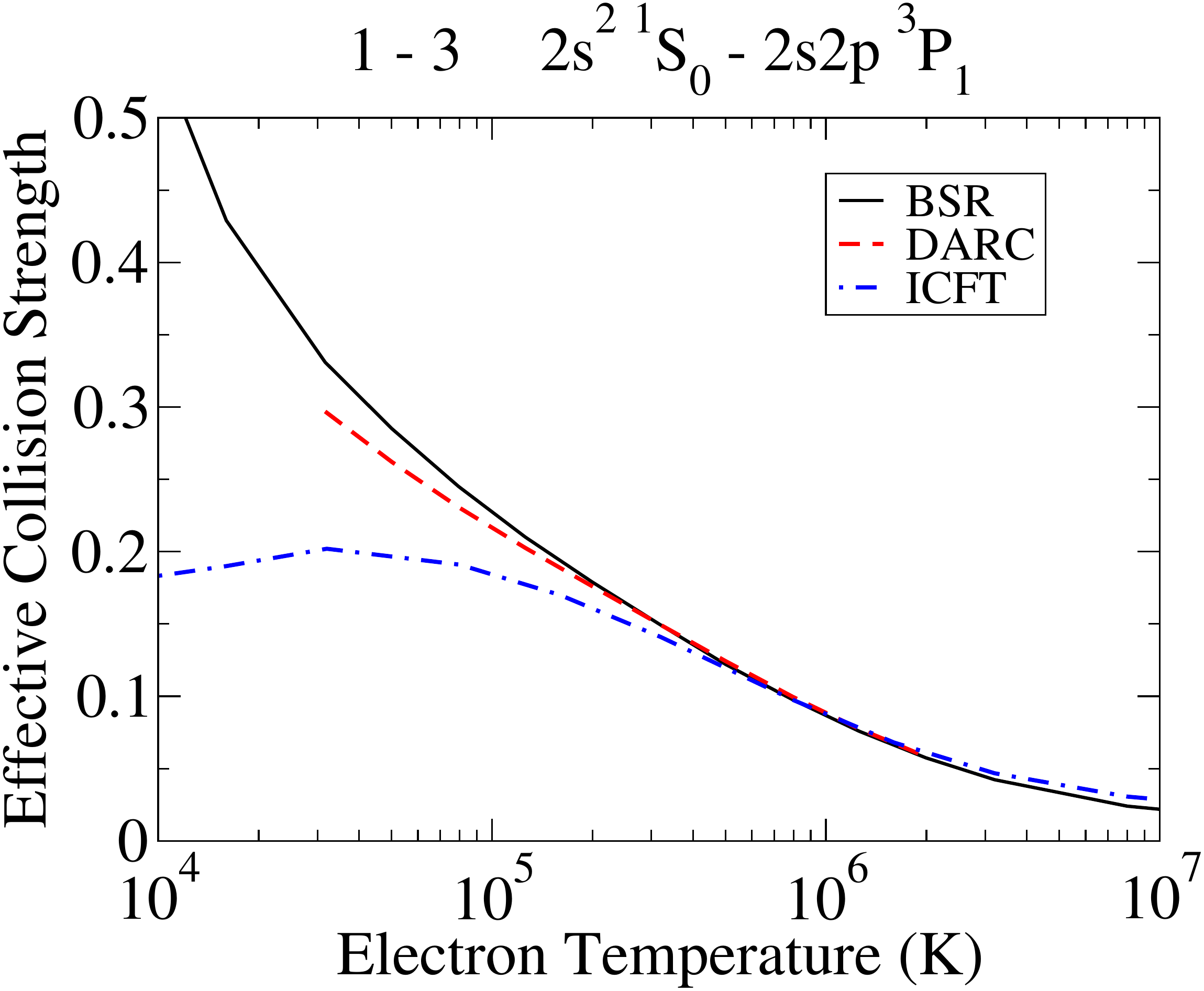}
   \includegraphics[width=0.38\textwidth,clip]{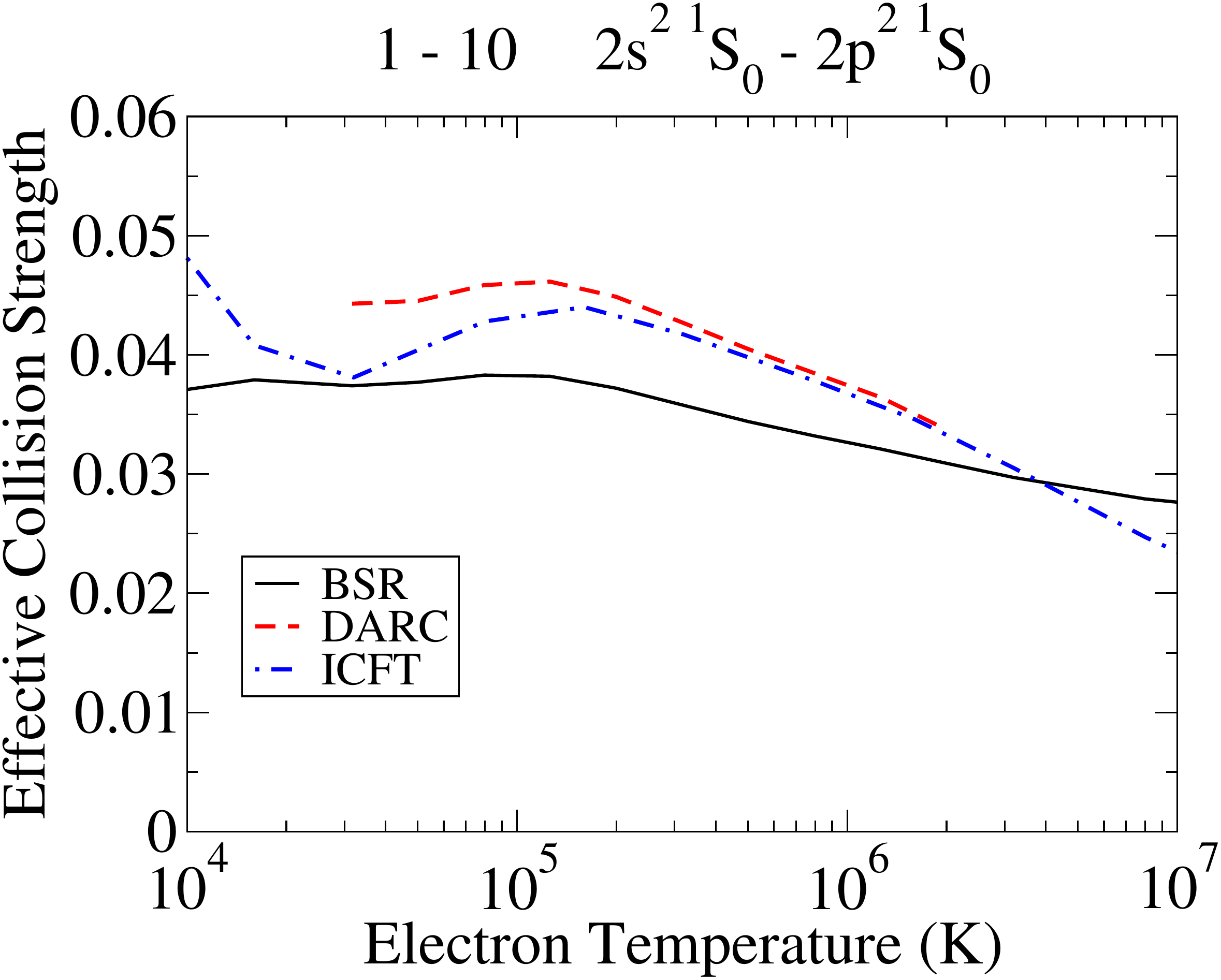} \includegraphics[width=0.38\textwidth,clip]{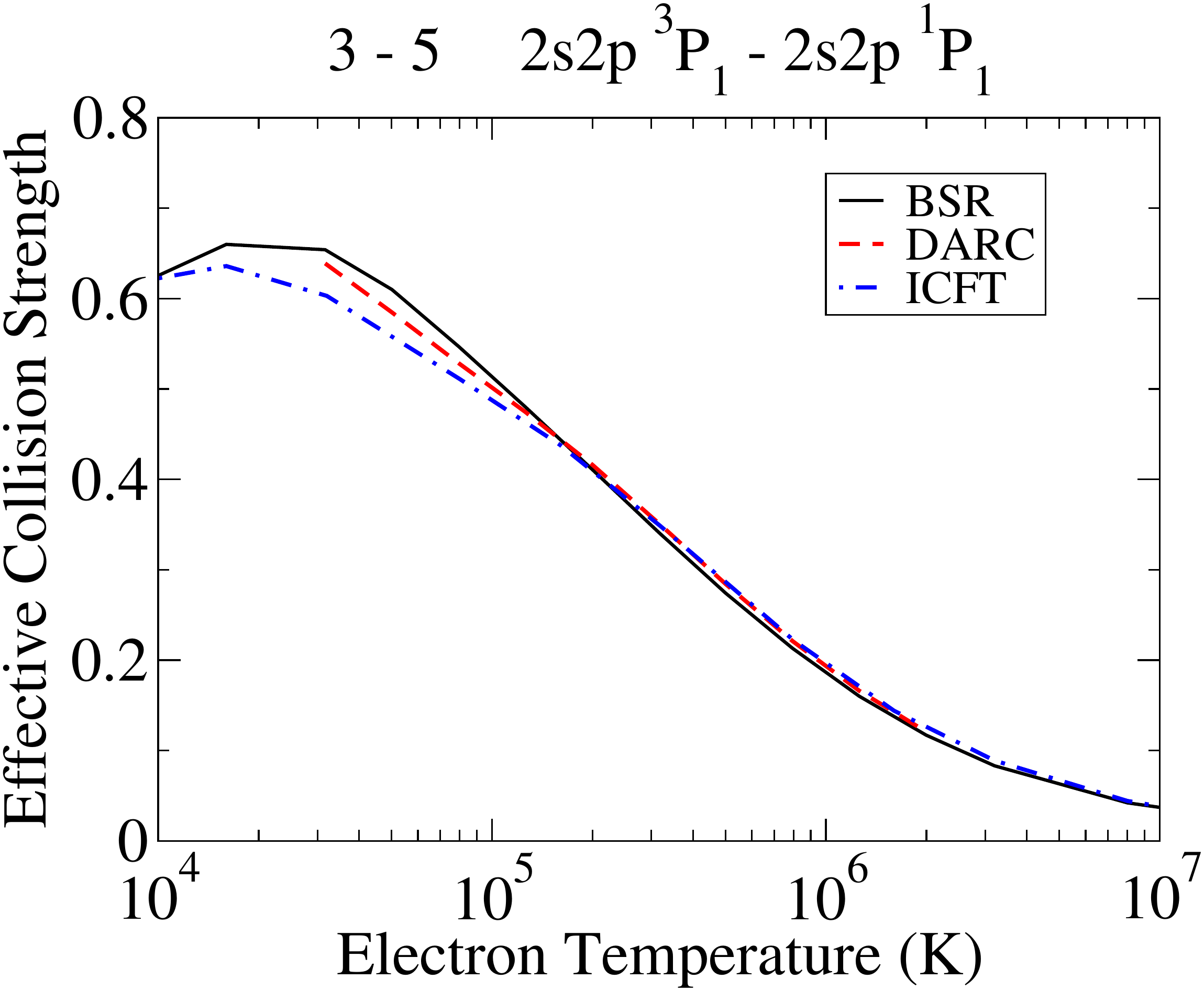}
\caption{Electron-impact excitation effective collision strengths versus electron temperature for $\mathrm{N}^{3+}$
         for some selected transitions. 
         Solid line: present work; 
         dashed line:~\cite{aggarwal2016};
         dash-dotted line:~\cite{fernandez-menchero2014a}.
      }
   \label{fig:n3ups}
\end{figure*}

Figure~\ref{fig:omega} exhibits the resonance structure of the
collision strength~$\Omega$ for three selected transitions,
one dipole-allowed with spin-change ($1\!-\!3$),
one M1$-$E2 with spin-change ($3\!-\!5$) and
finally a one-photon forbidden double-electron jump ($1\!-\!10$).
These results are from the present BSR
calculation and a previous ICFT model~\cite{fernandez-menchero2014a}.
DARC results for the same transitions are shown in~\cite{aggarwal2016}.
The resonance structure looks similar for these three transitions in BSR and ICFT,
while in the figures of~\cite{aggarwal2016} there are some prominent features
in the transitions $3\!-\!5$ and $1\!-\!10$ between $4.5$ and $5.5 \Ry$, respectively.
They are very narrow, however, and do not contribute significantly
to the effective collision strengths~$\Upsilon$, which are shown in Fig.~\ref{fig:n3ups}
along with the strong dipole transition $1\!-\!5$.
Results for the~$\Upsilon$s agree very closely,
to within $20\%$, over a wide range of temperatures,
except for the spin-changing transition $1\!-\!3$ $\mathrm{2s^2\,^1S_0 - 2s2p\,^3P_1}$
where~$\Upsilon$ decreases at low temperatures in the ICFT calculations.
This can be due to the position or the resolution of the resonances,
as will be discussed below.

\begin{figure*}
\centering
   \includegraphics[width=0.33\textwidth,clip]{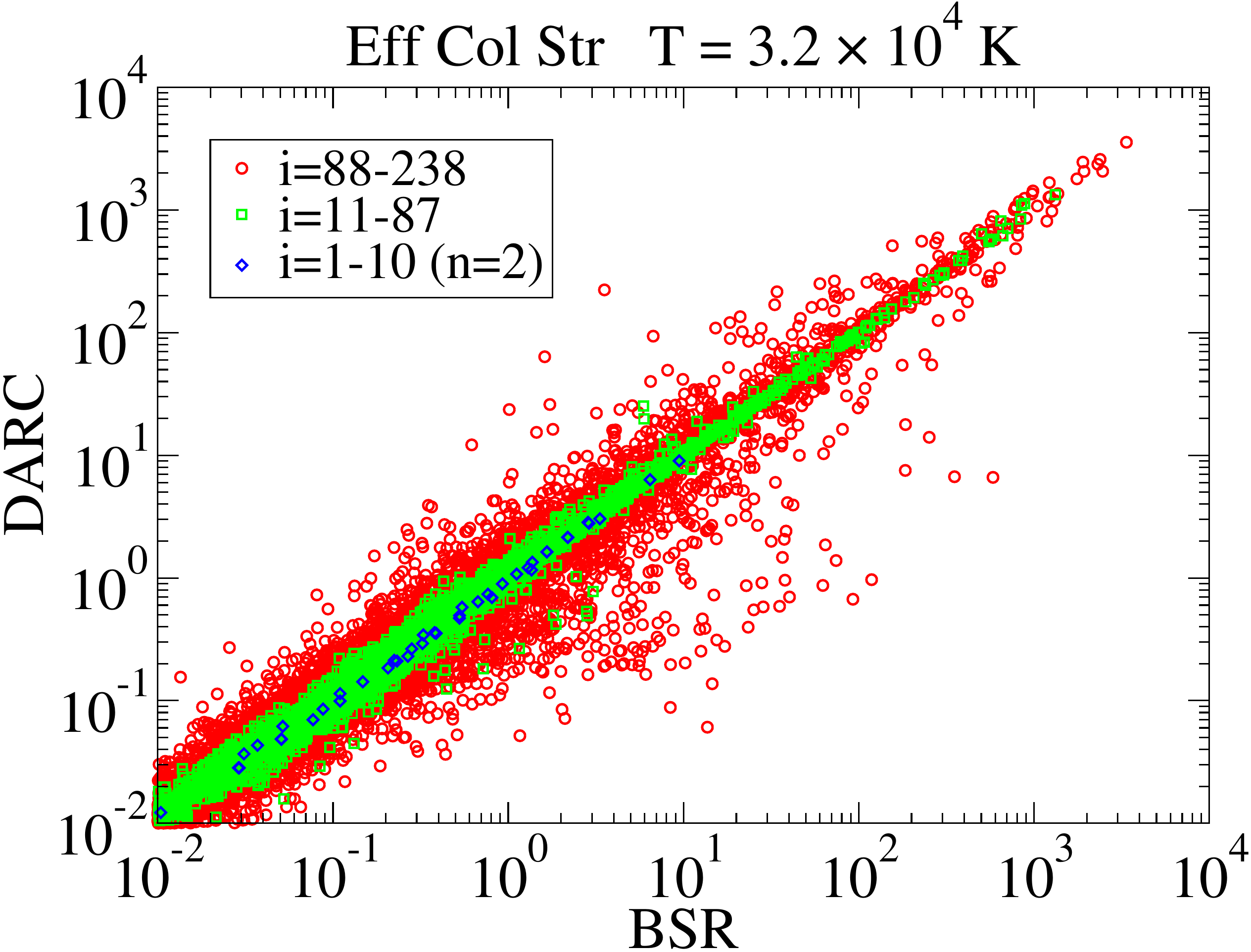} 
   \includegraphics[width=0.33\textwidth,clip]{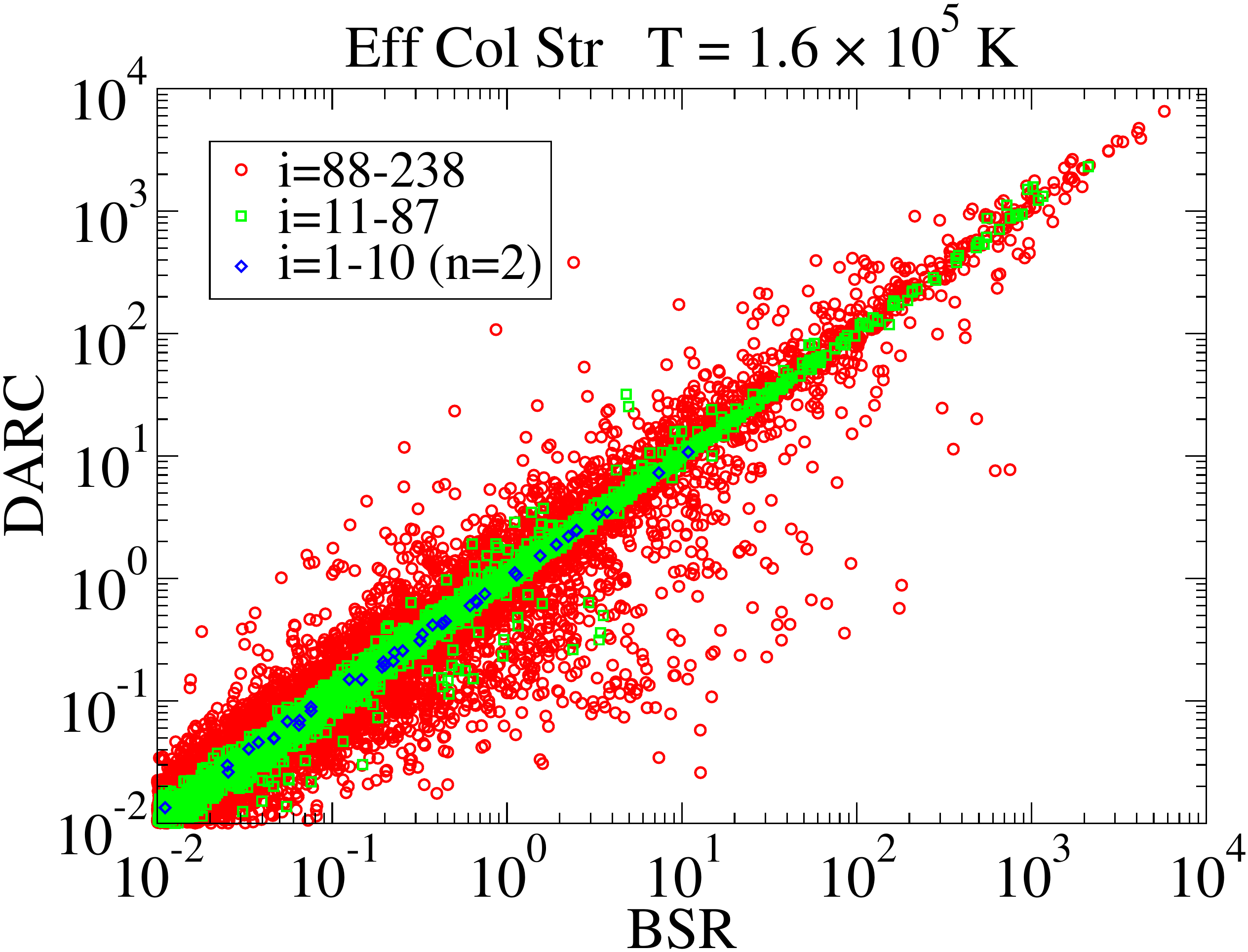}
   \includegraphics[width=0.33\textwidth,clip]{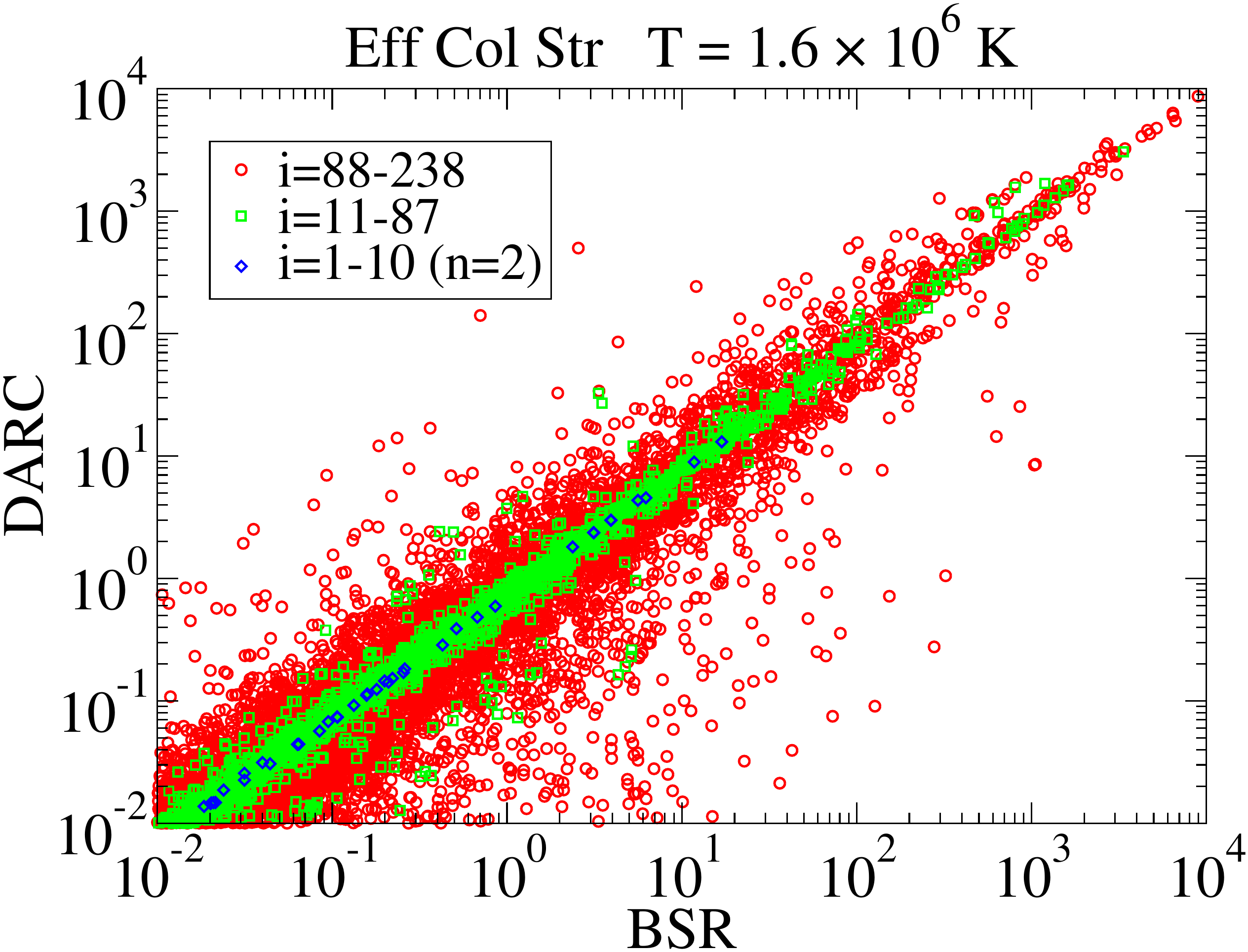}
   \includegraphics[width=0.33\textwidth,clip]{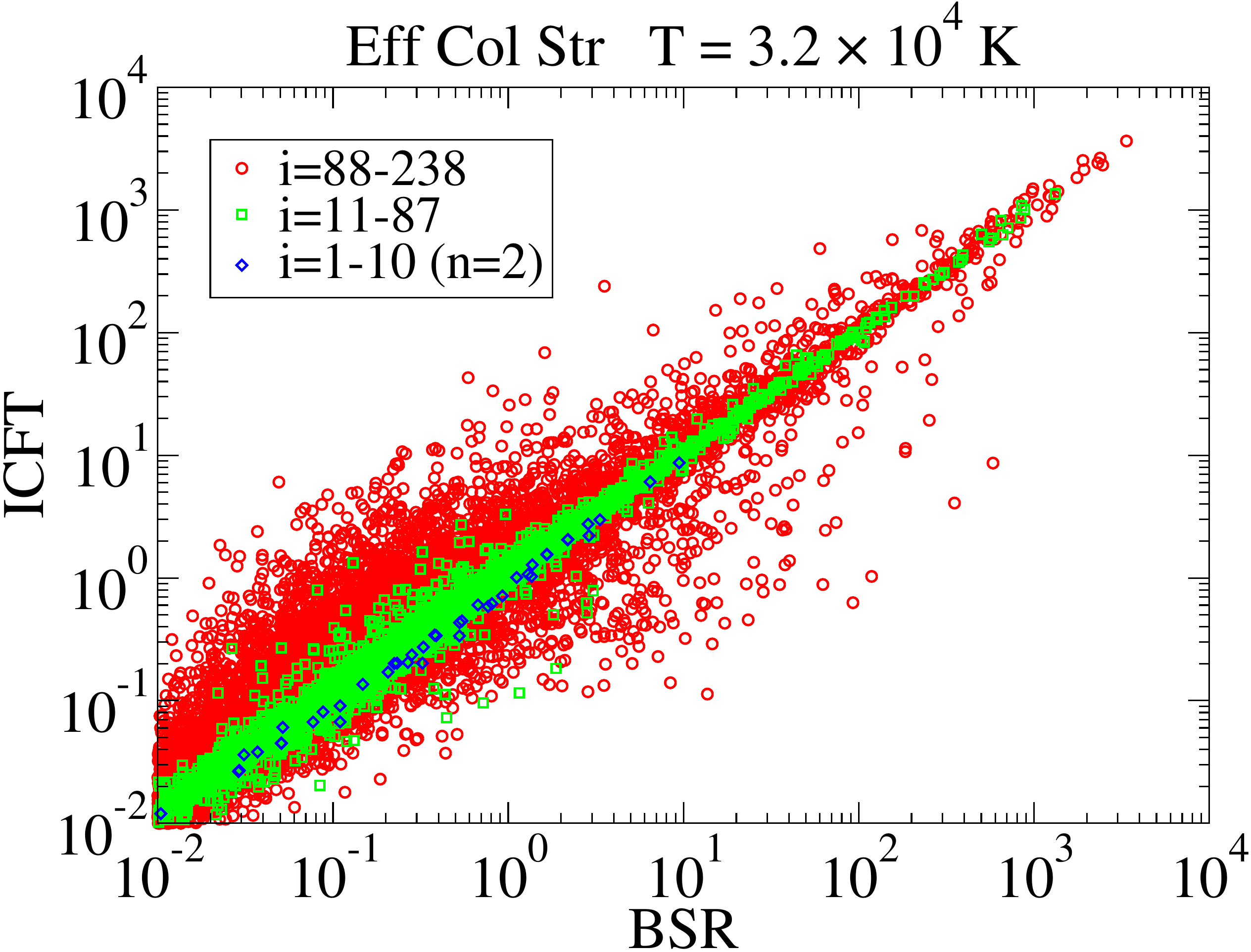}
   \includegraphics[width=0.33\textwidth,clip]{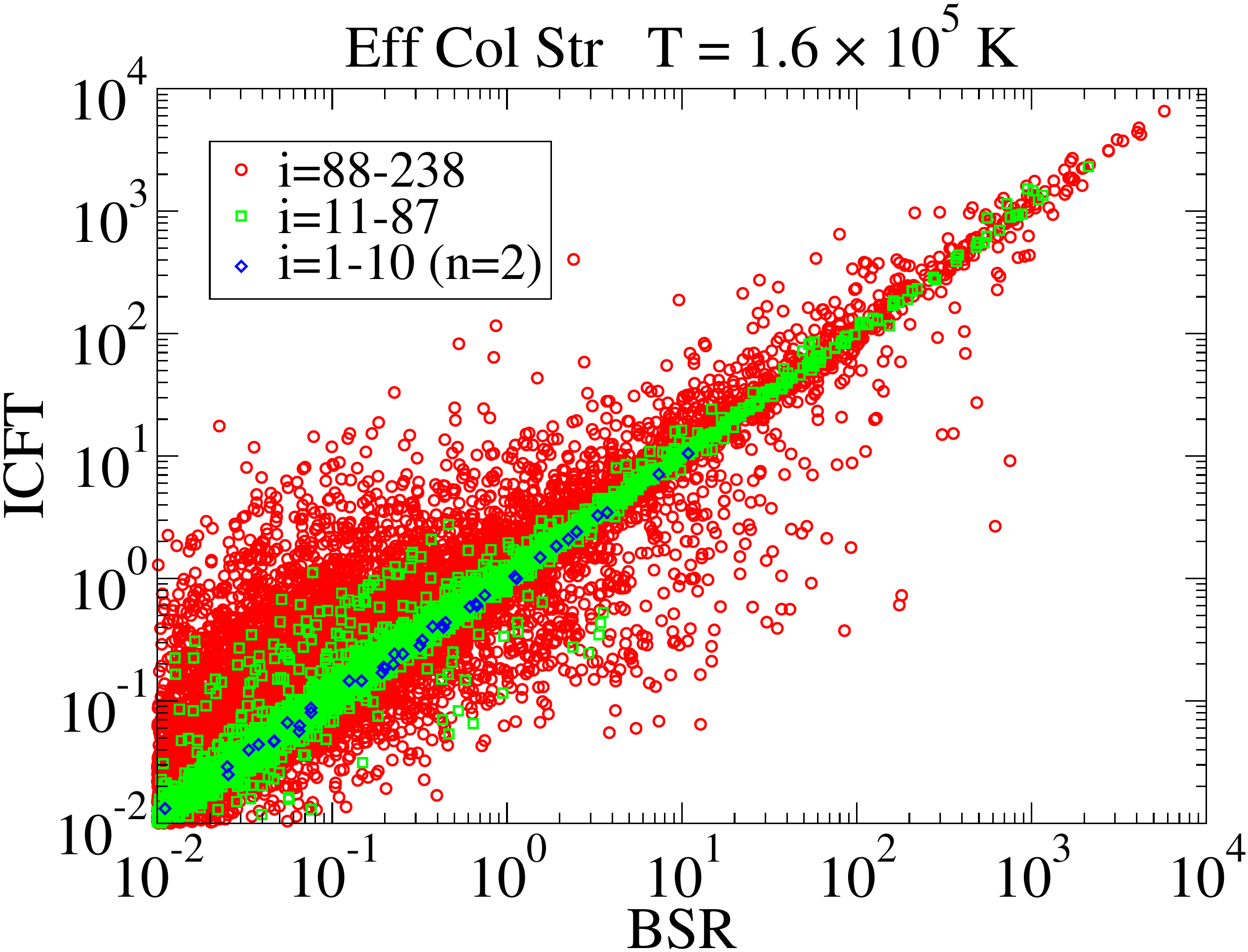}
   \includegraphics[width=0.33\textwidth,clip]{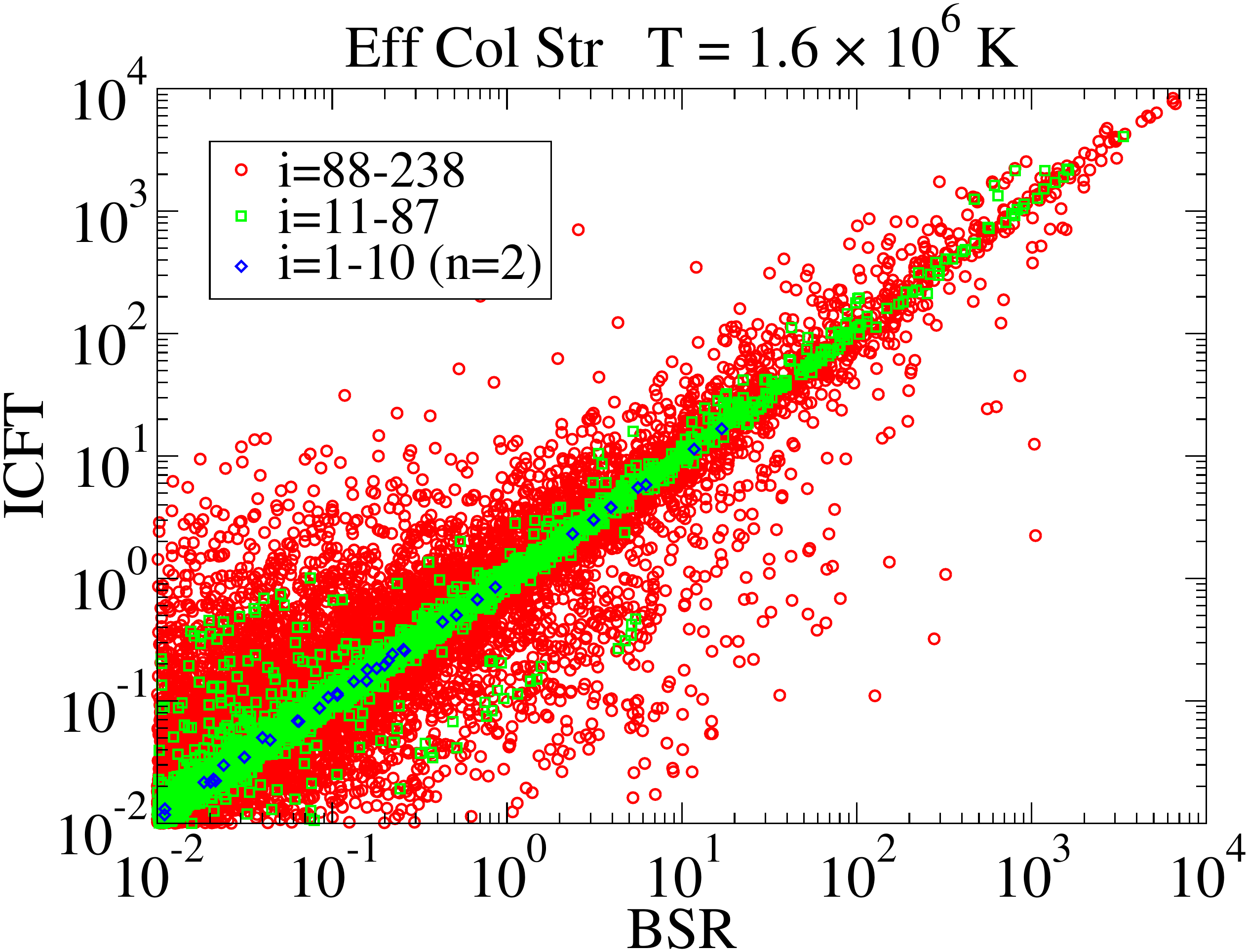}
   \caption{Comparison of effective collision strengths for the present BSR work with
      DARC~\cite{aggarwal2016} (upper panels) and
      ICFT~\cite{fernandez-menchero2014a} (lower panels) for
      three temperatures.
      Axes and symbols as in Fig.~\ref{fig:gfn3}.
      }
   \label{fig:n3upscomp}
\end{figure*}

Figure~\ref{fig:n3upscomp} shows a global comparison of the
effective collision strengths obtained 
previously~\cite{fernandez-menchero2014a,aggarwal2016} at three different
temperatures.
The dispersions at all three temperatures are similar to those
seen in the \hbox{$gf$-factors} in Fig.~\ref{fig:gfn3}.  This confirms
that the target structure is the principal source for the differences.
For further detail we display in Tables~\ref{tab:n3ups-BSR-DARC}
and~\ref{tab:n3ups-BSR-ICFT} the number of points in
Fig.~\ref{fig:gfn3} that deviate by more than a fixed percentage
from the diagonal.
In both comparisons, DARC or ICFT vs.\ BSR, the dispersions are
similar.  They are much smaller for transitions
with upper levels up to \#87 than for all levels.
This comparison, therefore, does not suggest which of the previous 
works~\cite{fernandez-menchero2014a,aggarwal2016} is of better quality.
We can only state that both are different and that the principal source for the
deviations originates in the target structure.

\begin{table}
\caption{Number of transitions in 
the upper panels of
Fig.~\ref{fig:n3upscomp} that differ by more than 
$\delta=|\Upsilon_{\mathrm{DARC}}-\Upsilon_{\mathrm{BSR}}|/\Upsilon_{\mathrm{BSR}}$.
}
\label{tab:n3ups-BSR-DARC}
\begin{indented}
\item[\hspace{8.0truemm}]\begin{tabular}{rrrrr}
   \br
   $\delta$ &  & \multicolumn{3}{c}{Temperature (K)} \\
   $(\%)$ & $gf$ & $3.2 \times 10^4$ & $1.6 \times 10^5$ & $1.6 \times 10^6$ \\
   \mr
   \multicolumn{5}{l}{All CC levels}     \\
   \mr
   10     & 6771 & 16644 & 16723 & 25681 \\
   20     & 5823 &  9313 & 10024 & 22964 \\
   50     & 4030 &  2377 &  3029 & 12076 \\
   100    & 1769 &   753 &  1033 &  4909 \\
   200    & 1342 &   348 &   538 &  2201 \\
   500    &  882 &   124 &   219 &   796 \\
   1000   &  583 &    68 &   113 &   411 \\
   \mr
   Total  & 8158 & 28203 & 28203 & 28203 \\
   \mr
   \multicolumn{5}{l}{CC levels $1-87$}  \\
   \mr
   10     &  698 &  1810 &  1523 &  3503 \\
   20     &  560 &   768 &   610 &  3205 \\
   50     &  320 &    79 &    68 &  1015 \\
   100    &  138 &    22 &    37 &   178 \\
   200    &  112 &    11 &    19 &    70 \\
   500    &   79 &     0 &     4 &    35 \\
   1000   &   49 &     0 &     0 &    12 \\
   \mr
   Total  & 1119 &  3741 &  3741 &  3741 \\
   \mr
   \multicolumn{5}{l}{Transition from levels $i=1-4$ to $j=1-87$}  \\
   \mr
   10     &   43 &   163 &    93 &   337 \\
   20     &   31 &    17 &     7 &   329 \\
   50     &   19 &     2 &     2 &    99 \\
   100    &   10 &     1 &     2 &     5 \\
   200    &    8 &     0 &     1 &     2 \\
   500    &    5 &     0 &     0 &     1 \\
   1000   &    2 &     0 &     0 &     1 \\
   \mr
   Total  &   90 &   338 &   338 &   338 \\
   \br
\end{tabular}
\end{indented}
\end{table}

\begin{table}
\caption{Number of transitions in
the lower panels of 
Fig.~\ref{fig:n3upscomp} that differ by more than 
$\delta=|\Upsilon_{\mathrm{ICFT}}-\Upsilon_{\mathrm{BSR}}|/\Upsilon_{\mathrm{BSR}}$.
}
\label{tab:n3ups-BSR-ICFT}
\begin{indented}
\item[\hspace{8.0truemm}]\begin{tabular}{rrrrr}
   \br
   $\delta$ &  & \multicolumn{3}{c}{Temperature (K)} \\
   $(\%)$ & $gf$ & $3.2 \times 10^4$ & $1.6 \times 10^5$ & $1.6 \times 10^6$ \\
   \mr
   \multicolumn{5}{l}{All CC levels}     \\
   \mr
   10     & 6788 & 20589 & 20839 & 21783 \\
   20     & 5900 & 14701 & 15402 & 17054 \\
   50     & 4055 &  5339 &  7195 &  9459 \\
   100    & 1782 &   571 &   736 &  1402 \\
   200    & 1340 &   317 &   460 &   883 \\
   500    &  910 &   127 &   203 &   462 \\
   1000   &  642 &    57 &    97 &   254 \\
   \mr
   Total  & 8158 & 28203 & 28203 & 28203 \\
   \mr
   \multicolumn{5}{l}{CC levels $1-87$}  \\
   \mr
   10     &  747 &  2215 &  2073 &  2206 \\
   20     &  576 &  1177 &  1020 &  1181 \\
   50     &  302 &   159 &   177 &   305 \\
   100    &  121 &    23 &    35 &    66 \\
   200    &   84 &    15 &    25 &    56 \\
   500    &   69 &     5 &    13 &    36 \\
   1000   &   58 &     0 &     1 &    14 \\
   \mr
   Total  & 1119 &  3741 &  3741 &  3741 \\
   \mr
   \multicolumn{5}{l}{Transition from levels $i=1-4$ to $j=1-87$}  \\
   \mr
   10     &   42 &   200 &   126 &   131 \\
   20     &   30 &    45 &    14 &    27 \\
   50     &   12 &     5 &     2 &     4 \\
   100    &    6 &     1 &     2 &     2 \\
   200    &    4 &     1 &     1 &     2 \\
   500    &    3 &     0 &     0 &     2 \\
   1000   &    2 &     0 &     0 &     0 \\
   \mr
   Total  &   90 &   338 &   338 &   338 \\
   \br
\end{tabular}
\end{indented}
\end{table}

As discussed in Sect.~\ref{sec:structure}, our target structure description appears to be 
significantly more accurate than those used in 
previous works.  
We cannot, however, state the same about the~$\Upsilon$ values.
Recall that we had to cut the CC expansion to a size for which
we could handle the resulting matrices.
It was shown in~\cite{fernandez-menchero2016a} that cutting
the CC expansion can lead to significant errors for transitions to
levels above those omitted from the CI expansion.
Consequently, we cannot properly assess the accuracy for
transitions involving level \#88 and above,
this time not because of limitations in the target structure
as in previous works but due to limitations in the CC
expansion.

\subsection{Peak abundance temperature}
\label{subsec:peakT}

The peak abundance temperature~$T_{\rm peak}$ of $\mathrm{N}^{3+}$ in collisional
plasmas lies around $126000 \kelvin$ \cite{mazzotta1998}.
At temperatures close to this, there is 
overall good agreement between the results from the three calculations for
transitions where the CI and CC expansions have essentially converged.
Tables~\ref{tab:n3ups-BSR-DARC} and~\ref{tab:n3ups-BSR-ICFT}
show that for levels below \#88, the predictions for approximately $80\%$ of the transitions
agree within $20\%$ for all calculations.

The deviations of the results obtained by the ICFT method are found mainly in
spin-changing transitions with small spin-orbit mixing in the target.
They might be caused by spin-orbit coupling of the colliding
electron with the target.
This is the only interaction that the ICFT formalism does not account for,
in contrast to the Breit-Pauli and the full-relativistic calculations.

\subsection{Low temperature}
\label{subsec:lowT}

In low-density photo-ionized plasmas, N$^{3+}$ may exist at temperatures much 
lower than~$T_{\rm peak}$.
In such a regime the most important contributions to $\Upsilon$
originate from resonances.
The positions of these resonances are directly connected with the energies of the target levels.
To generate good-quality results at low temperature, it is important to have the resonances 
in their correct position and well-resolved in energy.
Recall that the present calculation obtains the target energy levels closest to the NIST-recommended values.
It is also the one with the finest mesh in the resonance region $10^{-5} z^2$.
This suggests that the present work generated the most reliable results 
in the low-temperature regime.

The results for low temperatures could be improved by adjusting  the level energies 
to the recommended ones.
Such operation, however, perturbs the hamiltonian, including the non-diagonal 
matrix elements and, in turn,
may affect the results in the peak-abundance and high-temperature regimes, 
which are most important for many applications, specifically for collisional plasmas.
Usually the low-temperature region is treated individually in studies specially 
focused on this regime.
By analyzing our results in detail, we can confirm that the position of the resonances 
is the principal cause of disagreement at low temperatures between the BSR and ICFT results 
for the spin-changing transition $1-3$ shown in Fig.~\ref{fig:n3ups}.

\subsection{High temperature}
\label{subsec:highT}

The high-energy part of the collision strengths is mostly determined by the 
infinite energy point,
either the collision strength~$f$ for dipole-allowed transitions or the 
Born infinite-energy $\Omega_{\infty}^{\mathrm{B}}$ for the forbidden ones.
This point depends exclusively on the target structure, i.e., it is  
independent of the dynamical method used for the collision calculation, including
details of the close-coupling expansion.
Since the present model produces the most accurate target structure
for $\mathrm{N}^{3+}$ to date,
we expect it to also produce the most accurate results for 
electron-impact excitation of $\mathrm{N}^{3+}$ in the high-temperature region, 
including the highly-excited states above level~\#87.

\section{Conclusions}
\label{sec:conclusions}

The present work sheds new light on the important topic of effects  
due to the description of the target structure in collision calculations. 
We have independently obtained electron-impact excitation collision strengths
for the $\mathrm{N}^{3+}$ ion and carried out a thorough comparison with
predictions from previous ICFT and DARC calculations~\cite{fernandez-menchero2014a,aggarwal2016}.
The close-coupling expansion contained the same number of target states in  all calculations.
Comparing the results thus directly reveals the influence of the 
different structure descriptions. 

The large differences in the collision strengths from these calculations, seen in some
cases, suggest that 
the target structure is the main source of inaccuracy and uncertainty in these calculations, 
with the largest effects on transitions involving highly-excited levels.
The convergence of the CI expansion is essential for a proper description 
of the target, and 
in standard applications of the \hbox{R-matrix} method it can no longer be assessed as
one approaches the limit of the CC expansion.
An incomplete basis set leaves gaps in the spectrum.
This gap may contain several configurations or even entire Rydberg series, 
including the continuum.
When states in the CI expansion are strongly mixed and some are omitted, 
their effect on the scattering results can no longer be properly assessed.

Furthermore, one has to consider the usual limitation of the CC expansion in
the actual calculation, which is due to the available computational resources.
No close-coupling method can claim the validity of its results
for target states close to or even above the levels discarded in the CC 
expansion~\cite{fernandez-menchero2016a}. 

It has recently been emphasized that uncertainty estimates for theoretical 
predictions are essential for their usefulness in practical applications~\cite{chung2016}.
Even though this is a difficult task in the present work, we believe that the ICFT, DARC, and 
BSR results are accurate at the 20\% level for the majority of the strong transitions up to level \#87.
Nevertheless, the lack of coupling to the ionization continuum via a sufficient
number of pseudo-states in all the present models for the case of interest 
leaves room for surprises, especially for the weak transitions.
For a few levels, the differences in the various predictions reach a 
factor of two or even more. These differences are, once again, mainly due 
to the representation of the target structure.

The present work does not support the conjecture of 
Aggarwal {\em et al.}~\cite{aggarwal2016}  that the differences 
with the results of~\cite{fernandez-menchero2014a} are due to either the general validity  
or the practical implementation of the ICFT method.
The differences between the present BSR collision strengths and the corresponding ICFT results 
are approximately the same as those between the BSR and DARC calculations.
The only cases where the ICFT method may clearly lead to errors involves weak spin-changing
transitions between levels with small spin-orbit mixing,
which can be strongly influenced through the spin-orbit coupling with 
the incoming electron.
This mechanism is not considered by the ICFT method.
However, such weak transitions are in general not relevant 
for plasma modeling, certainly not in collisional plasmas.

Due to the superior target structure generated in the present work,
we believe that the present results are the currently best for 
electron collisions with~N$^{3+}$.
The differences between the BSR and the DARC / ICFT results may in fact serve as an uncertainty estimate
for the available excitation rates.
We are confident regarding the accuracy of the collision strengths for strong transitions among the
low-lying levels up \#87 in the $\mathrm{N}^{3+}$ spectrum, while 
the results for transitions to the higher-lying states remain questionable, especially
for the weak inter\-combination transitions. 
This situation is typical for many existing datasets and calls for further 
detailed consideration. 

\section*{Acknowledgments}

This work was supported by the United States National Science Foundation 
through grant No.\ PHY-1520970.  The numerical calculations were performed on STAMPEDE
at the Texas Advanced Computing Center.  They were made possible by the XSEDE allocation
No.\ PHY-090031.

\section*{References}
\providecommand{\newblock}{}

\end{document}